\documentclass{emulateapj}

\usepackage{lscape}
\usepackage{apjfonts}
\usepackage{amsmath}
\usepackage{graphicx}
\newcommand{\etal}{et al.}

\newcommand{\MgII}{Mg{\sevenrm II}}

 \font\sevenrm=cmr7 scaled 1000

\begin{document}
\title{Quasar Clustering from SDSS DR5: Dependences on Physical Properties}

\shorttitle{SDSS QUASAR CLUSTERING}

\shortauthors{SHEN ET AL.}

\author{Yue Shen\altaffilmark{1}, Michael A. Strauss\altaffilmark{1},
Nicholas P. Ross\altaffilmark{2}, Patrick B. Hall\altaffilmark{3},
Yen-Ting Lin\altaffilmark{1}, Gordon T. Richards\altaffilmark{4},
Donald P. Schneider\altaffilmark{2}, David H.
Weinberg\altaffilmark{5}, Andrew J. Connolly\altaffilmark{6},
Xiaohui Fan\altaffilmark{7}, Joseph F. Hennawi\altaffilmark{8},
Francesco Shankar\altaffilmark{5}, Daniel E. Vanden
Berk\altaffilmark{2}, Neta A. Bahcall\altaffilmark{1}, Robert J.
Brunner\altaffilmark{9}}

\altaffiltext{1}{Princeton University Observatory, Princeton, NJ
08544.}

\altaffiltext{2}{Department of Astronomy and Astrophysics, 525
Davey Laboratory, Pennsylvania State University, University Park,
PA 16802.}

\altaffiltext{3}{Dept. of Physics \& Astronomy, York University,
4700 Keele St., Toronto, ON, M3J 1P3, Canada.}

\altaffiltext{4}{Department of Physics, Drexel University, 3141
Chestnut Street, Philadelphia, PA 19104.}

\altaffiltext{5}{Astronomy Department, Ohio State University,
Columbus, OH 43210.}

\altaffiltext{6}{Department of Astronomy, University of
Washington, Box 351580, Seattle, WA 98195.}

\altaffiltext{7}{Steward Observatory, 933 North Cherry Avenue,
Tucson, AZ 85721.}

\altaffiltext{8}{Department of Astronomy, Campbell Hall,
University of California, Berkeley, California 94720.}

\altaffiltext{9}{Department of Astronomy, MC-221, University of
Illinois, 1002 West Green Street, Urbana, IL 61801.}


\begin{abstract}
Using a homogenous sample of $38,208$ quasars with a sky coverage
of $\sim 4000\,{\rm deg^2}$ drawn from the SDSS Data Release Five
quasar catalog, we study the dependence of quasar clustering on
luminosity, virial black hole mass, quasar color, and radio
loudness. At $z<2.5$, quasar clustering depends weakly on
luminosity and virial black hole mass, with typical uncertainty
levels $\sim 10\%$ for the measured correlation lengths. These
weak dependences are consistent with models in which substantial
scatter between quasar luminosity, virial black hole mass and the
host dark matter halo mass has diluted any clustering difference,
where halo mass is assumed to be the relevant quantity that best
correlates with clustering strength. However, the most luminous
and most massive quasars are more strongly clustered (at the $\sim
2\sigma$ level) than the remainder of the sample, which we
attribute to the rapid increase of the bias factor at the
high-mass end of host halos. We do not observe a strong dependence
of clustering strength on quasar colors within our sample. 
On the other hand, radio-loud quasars are more strongly clustered
than are radio-quiet quasars matched in redshift and optical
luminosity (or virial black hole mass), consistent with local
observations of radio galaxies and radio-loud type 2 AGN. Thus
radio-loud quasars reside in more massive and denser environments
in the biased halo clustering picture. Using the Sheth et al.
(2001) formula for the linear halo bias, the estimated host halo
mass for radio-loud quasars is $\sim 10^{13}\ h^{-1}M_\odot$,
compared to $\sim 2\times 10^{12}\ h^{-1}M_\odot$ for radio-quiet
quasar hosts at $z\sim 1.5$.
\end{abstract}

\section{INTRODUCTION}
\label{sec:intro}

Quasars are luminous, accreting supermassive black holes (SMBHs)
that are thought to reside at the center of almost every massive
galaxy (e.g., Salpeter 1964; Zel'dovich \& Novikov 1964;
Lynden-Bell 1969; Kormendy \& Richstone 1995; Richstone \etal\
1998). Despite their high luminosities and perhaps different BH
fuelling mechanism, quasars are no different from their
low-luminosity counterparts, active galactic nuclei (AGNs): they
both represent stages of building up SMBHs, and they interact with
the host galaxies in a self-regulated way (e.g., Silk \& Rees
1998; Magorrian et al. 1998; Ferrarese \& Merritt 2000; Gebhardt
et al. 2000; Kauffmann \& Haehnelt 2000; Wyithe \& Loeb 2002,
2003; King 2003; Di Matteo et al. 2005; Hopkins et al. 2006;
Croton et al. 2006). Therefore understanding the properties of
quasars/AGNs, such as their abundance (luminosity function),
spatial distributions (clustering), physical properties (spectral
energy distributions; black hole masses, etc.) and the dependence
of these properties on redshift, is crucial to understand galaxy
formation and SMBH growth within the standard hierarchical
structure formation framework.

In this work we focus on the spatial clustering of quasars.
Similar to galaxies, quasars can be used to trace the large-scale
distribution of the underlying dark matter. Because of their high
luminosity, quasars can be seen at redshift $0<z\lesssim6$ in a
single survey, allowing them to be used to map the dark matter
(DM) distribution over this entire redshift range. However, owing
to the enormously large spatial volume probed and the fact that
the quasar number density is much smaller than that of galaxies,
the analysis of quasar clustering has only become practical in
recent years, namely after large-scale surveys such as the 2dF
quasi-stellar object (QSO) redshift survey (2QZ, Croom \etal\
2004) and the Sloan Digital Sky Survey (SDSS, York \etal\ 2000).
Results on quasar clustering have been reported using the 2QZ
samples (e.g., Porciani, Magliocchetti \& Norberg 2004; Croom et
al. 2005; da \^{A}ngela et al. 2005; Porciani \& Norberg 2006),
the SDSS samples (e.g., Myers et al. 2006, 2007a, 2007b; Shen et
al. 2007a, 2008a; Ross et al. 2008), and the 2dF-SDSS LRG and QSO
survey (2SLAQ, Croom et al. 2008) sample (e.g., da \^{A}ngela et
al. 2008). Complementary to these results, quasar clustering has
also been studied through cross-correlations with large galaxy
samples (e.g., Adelberger \& Steidel 2005; Coil et al. 2007;
Padmanabhan et al. 2008), and surveys for close quasar pairs
(e.g., Hennawi et al. 2006; Myers et al. 2008).

While the statistics of quasar clustering are not yet comparable
to those of local galaxy clustering, recent measurements of the
quasar two-point correlation function (CF) already offer valuable
information about the physical properties of quasars, such as
their host dark matter halo masses and duty cycles. The amplitude
of the quasar correlation function suggests that quasars live in
massive dark matter halos, and therefore are biased tracers of the
underlying dark matter (e.g., Kaiser 1984; Bardeen et al. 1986);
the bias increases with increasing redshift (e.g., Croom et al.
2005; Myers et al. 2007a; Shen et al. 2007a; da \^{A}ngela et al.
2008; Ross et al. 2008). By comparing the relative abundance of
quasars and host halos, one can infer the average duty cycles of
quasar activity\footnote{Luminous quasars are sparse enough such
that most of the halos only host one quasar. Most of the multiple
quasars in single halos are missed in the SDSS spectroscopic
sample due to fiber collisions (see \S\ref{sec:data}), but the
fraction of such companion quasars gives a negligible contribution
to the quasar luminosity function as revealed in dedicated binary
quasar surveys (e.g., Hennawi et al. 2006; Myers et al. 2008).
Therefore no detailed halo occupation distribution modeling is
needed to connect the halo abundance to the quasar luminosity
function through the duty cycle. On the other hand, this simple
halo model assumes that halo mass is the only parameter that
determines the clustering properties of dark matter halos.
Recently it has been suggested that halo clustering also depends
on the assembly history and recent merger activity (e.g., Gao et
al. 2005; Wechsler et al. 2006; Furlanetto \& Kamionkowski 2006;
Wetzel et al. 2007); such effects will complicate the
interpretation of quasar clustering measurements (e.g., Wyithe \&
Loeb 2008). Nevertheless we neglect these complications in the
current work, since the magnitude of such effects is still quite
uncertain at this moment.} (e.g., Cole \& Kaiser 1989; Martini \&
Weinberg 2001; Haiman \& Hui 2001). With substantial
uncertainties, the estimated host DM halo mass is $\sim
10^{12-13}\ M_\odot$ (e.g., Porciani et al. 2004; Croom et al.
2005; Myers et al. 2006, 2007a; Shen et al. 2007a; da \^{A}ngela
et al. 2008; Padmanabhan et al. 2008), and the estimated quasar
lifetime is between a few Myr and $10^8$ yr. Quasar clustering
measurements have also facilitated several theoretical
investigations on the cosmic evolution of SMBHs within the
hierarchical structure formation paradigm (e.g., Hopkins et al.
2007; Shankar et al. 2008a; White et al. 2008; Croton 2008;
Shankar et al. 2008b; Wyithe \& Loeb 2008).

Previous investigations also examined the potential luminosity
dependence of quasar clustering. If quasar luminosity is a good
indicator of the host halo mass, then more luminous quasars live
in more massive halos, and hence should have stronger clustering.
However, the current observations only show weak luminosity
dependence of quasar clustering at $z\lesssim 2.2$ (e.g.,
Adelberger \& Steidel 2005; Croom et al. 2005; Porciani \& Norberg
2006; Myers et al. 2007a; da \^{A}ngela et al. 2008), albeit with
large error bars and, in most cases, limited dynamical range
($\lesssim 1$ dex in luminosity). If indeed quasar clustering
depends only weakly on luminosity, then low luminosity quasars can
also live in high mass halos and vice versa, which means there is
substantial scatter in the host halo mass--quasar luminosity
relation. This is naturally expected in some evolutionary quasar
models, where quasar luminosity also depends on the evolving
accretion rate onto the SMBH (e.g., Hopkins et al. 2005; Hopkins
et al. 2006; Shen et al. 2007b). Based on numerical simulations of
quasar light curves and prescriptions for the relation between
quasar peak luminosity and host halo mass, Lidz et al. (2006) were
able to show a weak luminosity dependence of quasar clustering at
$z\sim 2$ for intermediate luminosity ranges. They argue that the
peak luminosity is a better tracer of the host halo mass than the
instantaneous luminosity, although in practice one can only
observe the instantaneous luminosity.

An alternative approach is to study quasar clustering as a
function of black hole mass, which might better correlate with
host halo mass than luminosity (e.g., Ferrarese 2002; Baes et al.
2003). Unfortunately for most quasars, the only estimates of their
BH mass are virial BH masses (e.g., McLure \& Jarvis 2002;
Vestergaard \& Peterson 2006; Shen et al.\ 2008c), which are based
on broad line widths and continuum luminosities measured from
single-epoch spectra. Substantial uncertainties and biases exist
for these estimates (e.g., Shen et al. 2008b), and any potential
difference in clustering may be washed out by these complications
(see the discussion in \S\ref{subsec:BH_vir_vs_Mhalo}).
Nevertheless, we will use these virial BH mass estimates to test
any dependence of quasar clustering on BH mass, which in turn can
offer some insight on the accuracy of these virial mass estimates.

There are also sub-classes of quasars that show characteristic
properties different from the ``regular'' quasar population.
Quasars have a distribution of colors at a given redshift. Using a
sample of 4576 uniformly-selected quasars from SDSS, Richards et
al. (2003) found that at each redshift the $g-i$ color follows an
approximate Gaussian distribution with a red tail, and they
proposed to use the deviation from the median $g-i$ color as a
redshift-independent color indicator. Thus quasars can be divided
into blue, red and reddened populations (e.g., Richards et al.
2003), where the reddened population is most likely dust-reddened
in the quasar host galaxy (e.g., Hopkins et al. 2004). There are
also significant differences (at least in the ensemble average)
between the emission line properties for blue and red quasars
(e.g., Richards et al. 2003). Moreover, a small fraction
($\lesssim 10\%$) of quasars are radio-loud, although they have
similar optical properties to radio-quiet quasars. It will be
interesting to study their clustering properties for these quasar
sub-classes to see if their large-scale environment is responsible
for their different properties.

The purpose of this paper is to investigate various dependences of
quasar clustering using a homogeneous spectroscopic sample drawn
from the SDSS Fifth Data Release (DR5, Adelman-McCarthy et al.
2007). In particular we focus on the dependences on luminosity,
virial BH mass, quasar color and radio loudness. The redshift
evolution of quasar clustering/bias factor and redshift
distortions at $z\le 2.2$ are reported in a companion paper (Ross
et al. 2008). In \S\ref{sec:data} we describe our sample. We
present clustering measurements in \S\ref{sec:CF} and discuss our
results in \S\ref{sec:diss}. We adopt a flat $\Lambda$CDM
cosmology with cosmological parameters consistent with the three
year WMAP observations (Spergel et al. 2007): $\Omega_M=0.26$,
$\Omega_{\Lambda}=0.74$, $h=0.7$, $\Omega_b=0.0435$, $n_s=0.95$,
$\sigma_8=0.78$. Comoving distance will be used throughout unless
otherwise specified.

\section{DATA}\label{sec:data}

The SDSS uses a dedicated 2.5-m wide-field telescope (Gunn \etal\
2006) with a drift-scan camera with 30 $2048 \times 2048$ CCDs
(Gunn \etal\ 1998) to image the sky in five broad bands
($u\,g\,r\,i\,z$; Fukugita \etal\ 1996).  The imaging data are
taken on dark photometric nights of good seeing (Hogg \etal\
2001), are calibrated photometrically (Smith \etal\ 2002; Ivezi\'c
\etal\ 2004; Tucker \etal\ 2006) and astrometrically (Pier \etal\
2003), and object parameters are measured (Lupton \etal\ 2001;
Stoughton \etal\ 2002). Quasar candidates for spectroscopy are
selected from the imaging data using their colors (Richards \etal\
2002), and are arranged in spectroscopic plates (Blanton \etal\
2003) to be observed and confirmed with a pair of double
spectrographs.

The SDSS spectroscopy prohibits two targets closer than $55''$ to
be assigned spectroscopic fibers on one plate due to fiber
collision. Although serendipitous close pairs can be observed on
overlap plates (Hennawi et al. 2006), their completeness is poor
in the SDSS main quasar sample. Fig. 1 in Ross et al. (2008) shows
the minimal projected comoving separations corresponding to the
$55''$ fiber collision scale as function of redshift. These are
the smallest scales that can be systematically probed by the SDSS
spectroscopic quasar sample at each redshift. Investigations on
quasar clustering below these scales can only be achieved with
other dedicated quasar samples (e.g., Hennawi et al. 2006; Myers
et al. 2008).

\subsection{Sample Selection}\label{subsec:sample}

\begin{figure*}
  \centering
    \includegraphics[width=0.45\textheight]{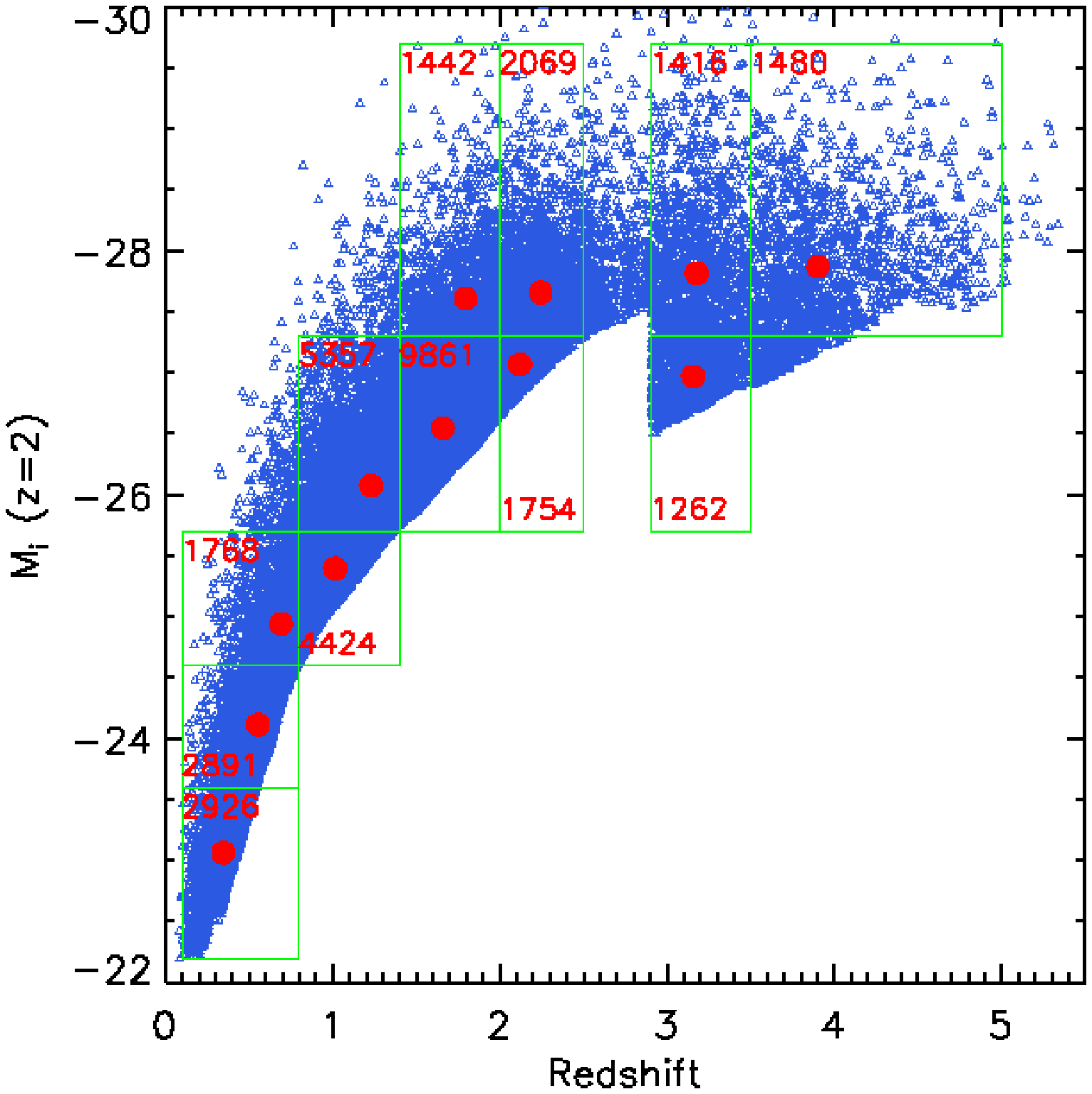}
    \includegraphics[width=0.45\textheight]{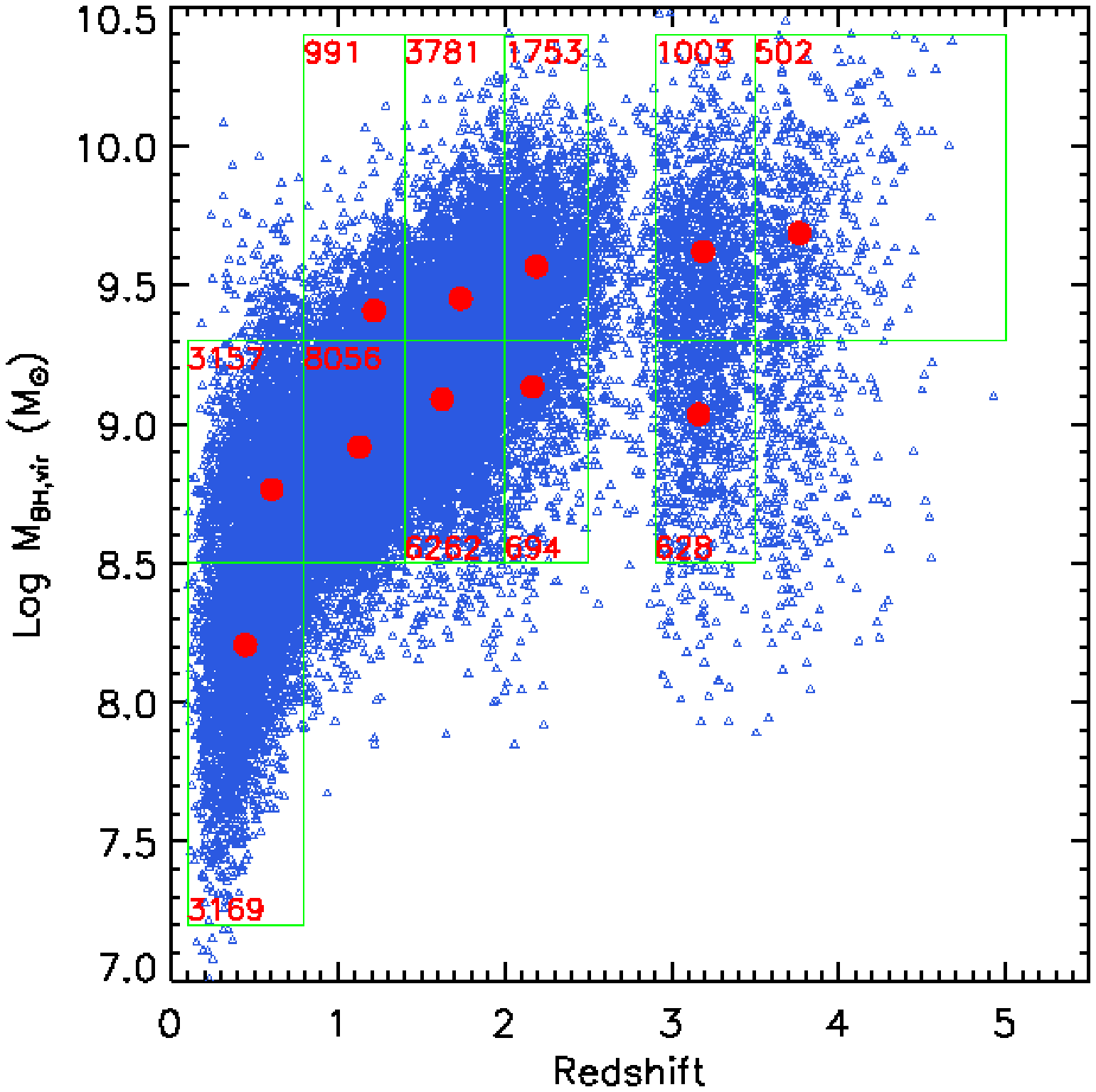}
    \caption{Distributions of quasars in the redshift-luminosity (left) and redshift-virial mass (right)
     planes, for our uniform quasar sample. The superposed grids indicate the subsamples we use to explore
     dependences of clustering strength on luminosity and virial BH mass. In each bin,
    the number of quasars is marked, and the red filled circle marks the location
of the median value. Note that these grids avoid straddling the
redshift deserts at $z\sim 2.7$ and $z\sim 3.5$ due to quasar
selection inefficiency (see text).}
    \label{fig:z_L_BH_div}
\end{figure*}

We take the DR5 spectroscopic quasar catalog (Schneider et al.
2007) as our parent sample from which we draw our clustering
samples. This catalog contains 77,429 bona fide quasars that have
luminosities larger than $M_i=-22.0$ (using a slightly different
cosmology in that paper) and have at least one broad emission line
(${\rm FWHM}>1000 {\rm\ km\ s^{-1}}$) or have interesting/complex
absorption features. About half of the quasars in this catalog
were selected from a uniform algorithm (as described in Richards
\etal\ 2002) implemented after DR1 (Abazajian et al. 2003), which
is flux limited to\footnote{All magnitudes in this paper are
corrected for Galactic extinction using the Schlegel, Finkbeiner
\& Davis (1998) map.} $i=19.1$ at $z\lesssim 3$ and $i=20.2$ at
$z\gtrsim 3$. There are a few ($\lesssim 2\%$) $i>19.1$ quasars at
$z\lesssim 3$ which were selected by the high-$z$ ($griz$) branch
of the targeting algorithm (Richards \etal\ 2002); we reject these
objects when constructing the homogenous clustering sample. Our
final clustering sample thus includes all uniformly-selected
quasars with limiting magnitude $i=19.1$ at $z< 2.9$ and $i=20.2$
at $z\ge 2.9$. This homogeneous clustering sample includes 38,208
quasars covering the redshift range $0.1\lesssim z\lesssim 5.3$;
the angular coverage of this sample is identical to the high-$z$
($z\ge 2.9$) quasar sample studied in Shen et al. (2007a, the {\em
all} sample), and covers a solid angle of $\sim 4000$ deg$^2$. We
do not consider further refinements of the sample, such as using
only {\em good} imaging fields as in Richards et al. (2006) and
Shen et al. (2007a), because at low $z$, quasars are found by UV
excess, and are not as subject to subtle problems in the
photometry as are high-$z$ quasars, and we generally found
consistent results with and without including bad fields (e.g.,
see appendix C in Ross et al. 2008). Although we may gain better
control over systematics, such refinements will reduce the number
of quasars in each subsample and therefore decrease the S/N for
clustering measurements. The entire clustering sample is shown in
the left panel of Fig. \ref{fig:z_L_BH_div} in the space of
redshift and $i-$band absolute magnitude ($K-$corrected to $z=2$,
see below and Richards et al. 2006). This homogeneous quasar
sample is also used in the companion paper (Ross et al. 2008),
where it is referred to as the UNIFORM sample.

We divide our sample into six redshift bins: $[0.1,0.8]$,
$[0.8,1.4]$, $[1.4,2.0]$, $[2.0,2.5]$, $[2.9,3.5]$, and
$[3.5,5.0]$, where we have deliberately avoided straddling two
redshift deserts of quasars at $z\sim 2.7$ and $z\sim 3.5$, caused
by the inefficiency of quasar color selection at these redshifts
due to the contamination from F and GK stars (Fan 1999; Richards
et al. 2002; see also fig. 1 in Richards et al. 2006). Note that
these redshift bins are different from those used in Ross et al.
(2008). While the redshift evolution of quasar clustering is not
the focus of this paper, we measure quasar clustering for these
redshift bins independently from Ross et al. (2008) as a
consistency check. Within each redshift bin, we further divide the
sample according to luminosity or virial black hole mass, where we
take the virial mass estimates from Shen et al. (2008b). These
redshift grids are shown in Fig. \ref{fig:z_L_BH_div} for $L-z$
(left) and $M_{\rm BH,vir}-z$ (right). We use the $K-$corrected
absolute $i-$band magnitude $M_i(z=2)$ normalized at $z=2$ (e.g.,
Richards et al. 2006) as a luminosity indicator. The conversion
between this non-standard absolute magnitude and $M_i(z=0)$ is
given by equation (1) in Richards et al. (2006). For a universal
power-law spectral energy distribution (SED) with
$\alpha_\nu=-0.5$ we simply have $M_i(z=0)=M_i(z=2)+0.5964$. There
is also a tight correlation between $M_i(z=2)$ and the bolometric
luminosity $L_{\rm bol}$ estimated from the quasar spectrum, as
shown in fig. 2a of Shen et al. (2008b). This correlation is
described by\footnote{A fixed slope of $b=-2.5$ is used in fitting
$M_i(z=2)=a+b\log(L_{\rm bol}/{\rm erg s^{-1}})$. If we do not fix
$b$ then the best fitted parameters are $a=94\pm 11$,
$b=-2.58\pm0.23$, which is fully consistent with a nominal slope
$b=-2.5$ between luminosity and magnitude.}:
\begin{equation}\label{eqn:Mi_Lbol}
M_{i}(z=2)=90-2.5\log (L_{\rm bol}/{\rm erg s^{-1}})\ ,
\end{equation}
where the Gaussian scatter of the correlation is $\sigma=0.166$.

While the $L-z$ and $M_{\rm BH,vir}-z$ grids in Fig.
\ref{fig:z_L_BH_div} can be used to study the luminosity
dependence and virial mass dependence at fixed redshift, the small
number of objects in each bin will limit the quality of clustering
measurements. We thus construct four other divisions using a
larger redshift range $0.4\le z\le 2.5$ to increase the
signal-to-noise ratio, which are the following:

(a) {\em median luminosity (bright and faint samples).} The sample
is divided by the median luminosity at each redshift, which is
used to test luminosity dependence. In addition we also divide the
sample into the top $10\%$ luminous quasars and the remaining less
luminous quasars, since if these brightest quasars live in the
rarest halos then their clustering will be much stronger than the
less luminous ones because of the rapid increase of linear bias
factor with halo mass (e.g., Kaiser 1984; Mo \& White 1996).

(b) {\em median virial mass (high and low mass samples).} The
sample is divided by the median virial BH mass at each redshift,
which is used to test virial BH mass dependence. As in the
luminosity case, we also divide the sample into the top $10\%$
massive quasars and the remainder.

(c) {\em color division (reddened, red and blue samples).} The
sample is divided according to {\em relative} quasar colors. We
use $\Delta(g-i)$, i.e., the deviation from the median $g-i$ color
at each redshift for our sample\footnote{Note these $\Delta(g-i)$
values are re-calculated for our uniformly-selected sample, not
those tabulated in the Schneider et al. (2007) catalog.}, as a
color indicator (described in detail in Richards et al. 2003). The
reddened quasars are classified as those with
$\Delta(g-i)>1.5\sigma$ from the median value $\Delta(g-i)=0$,
where $\sigma$ is the standard deviation of the distribution of
$\Delta(g-i)$; the red quasars are classified as those with
$0\le\Delta(g-i)\le 1.5\sigma$; the blue quasars are those with
$\Delta(g-i)<0$. These subsamples are used to test color
dependence.

(d) {\em radio division (radio and non-radio samples).} The sample
is divided into FIRST-detected and undetected sources using the
target selection information included in the DR5 quasar catalog.
The FIRST-detected and undetected quasars within this redshift
range $0.4\le z\le 2.5$ are complete to $i=19.1$ in our uniform
sample. The FIRST-detected sample contains quasars that are
generally radio-loud with typical radio-loudness ${\cal R}\equiv
f_{\rm 6\ cm}/f_{2500}\sim 100$ (Jiang et al. 2007). These
subsamples are used to test radio dependence. To simplify the
discussion we use {\em FIRST-detected/undetected} and {\em
radio-loud/quiet} interchangeably in the following sections, but
we acknowledge that they are two different concepts. Note that the
flux limit of the FIRST survey is bright enough that moderately
radio-loud objects ($R\sim 10$) may be undetected.

\begin{figure*}
  \centering
    \includegraphics[width=0.45\textwidth]{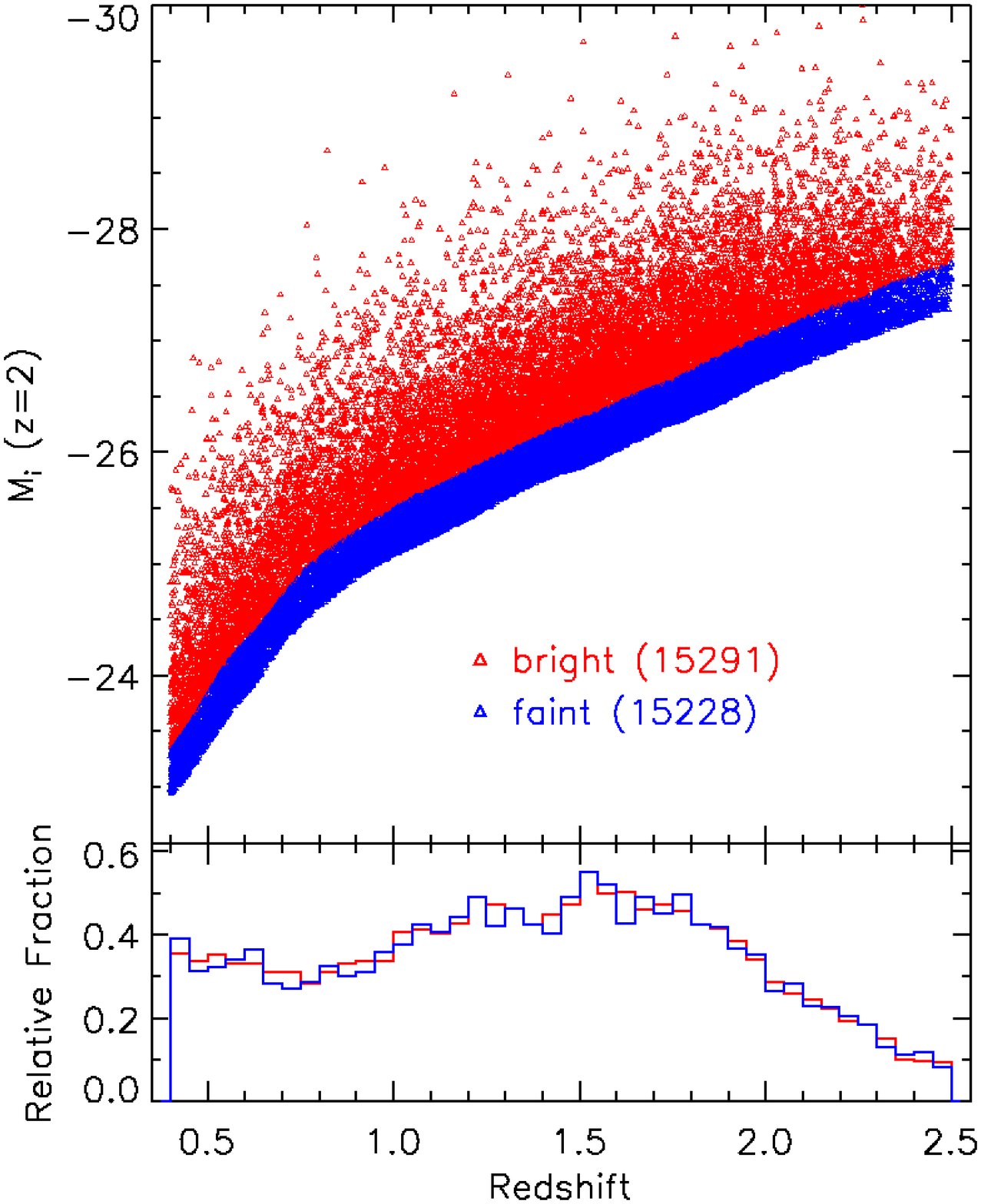}
    \includegraphics[width=0.45\textwidth]{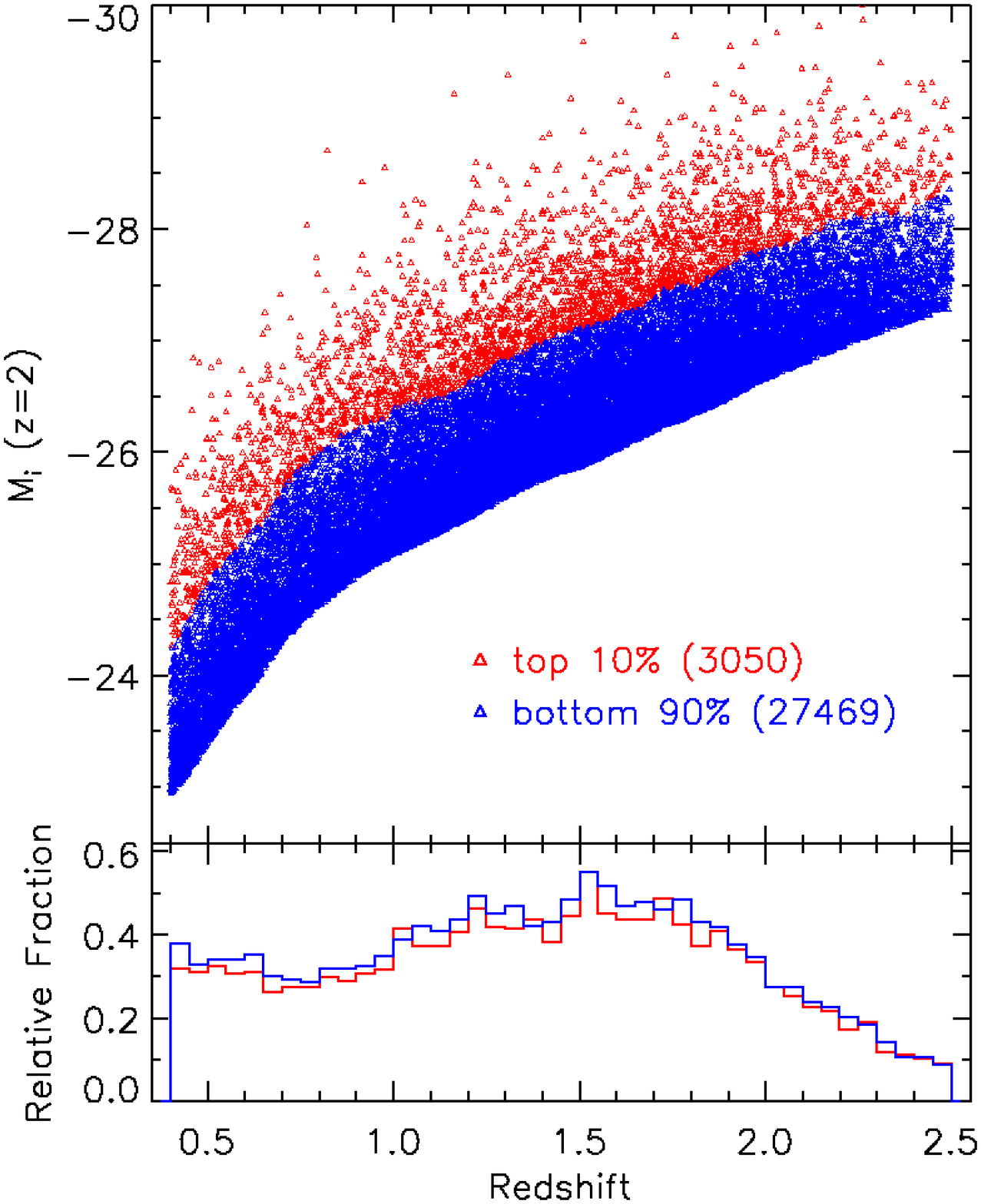}
    \includegraphics[width=0.45\textwidth]{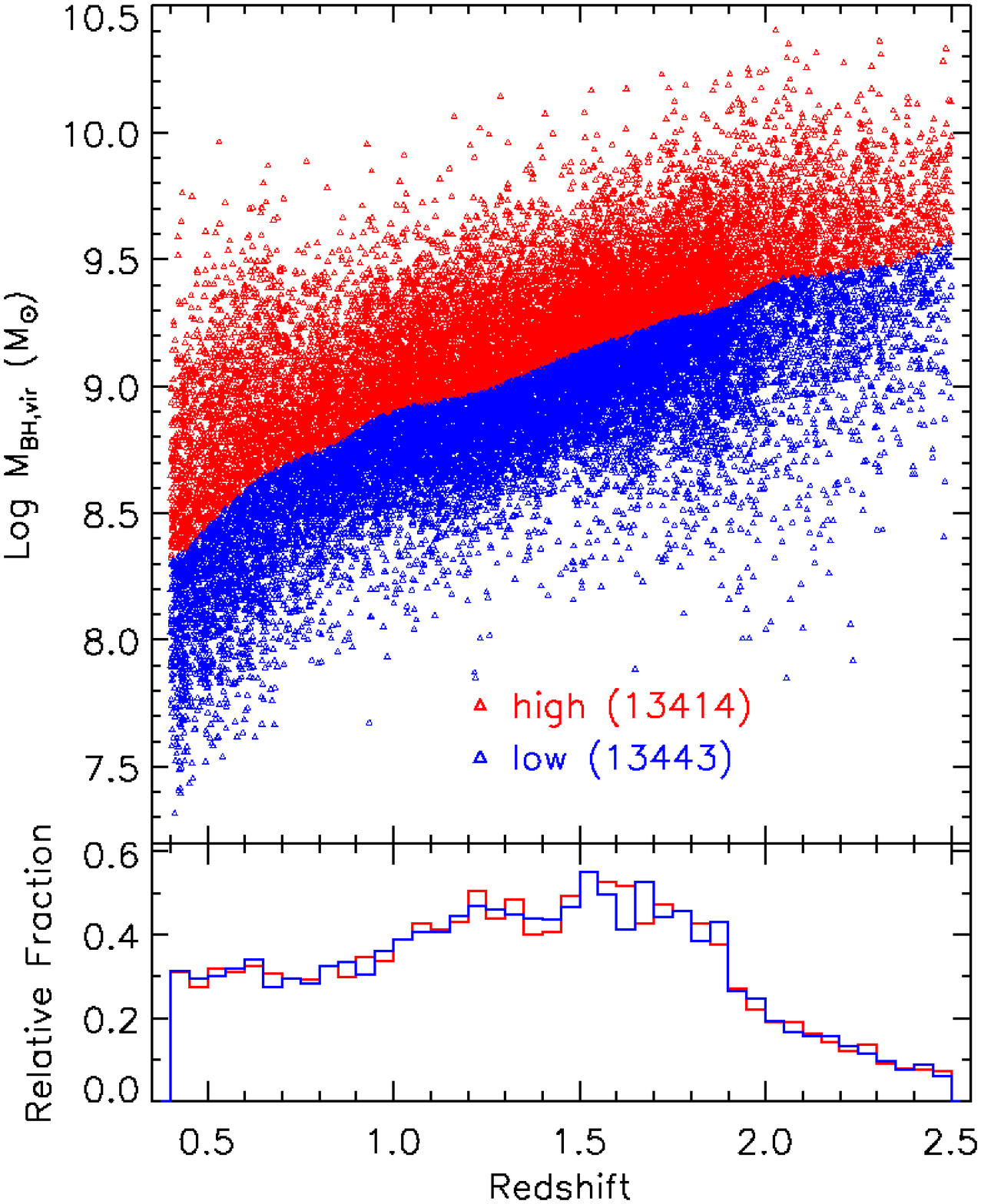}
    \includegraphics[width=0.45\textwidth]{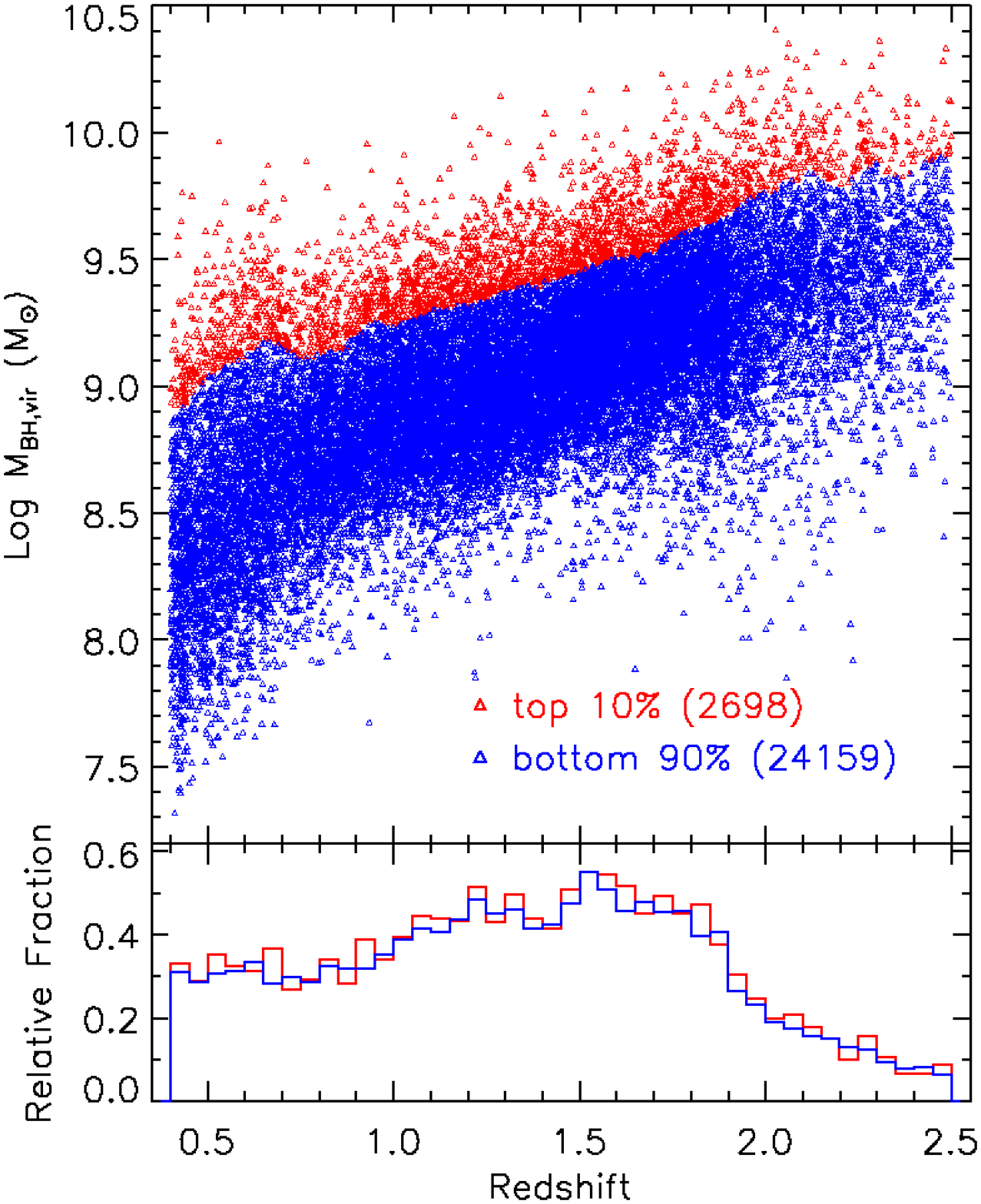}
    \caption{Sample divisions at median and $10\%$ highest luminosities (upper panels) and virial masses (bottom panels), in the redshift
    range $0.4\le z\le 2.5$. In the bottom of each panel we show the normalized redshift distributions
    for each subsample. The similarity of their redshift distributions allows a fair comparison
    of their relative clustering strength.}
    \label{fig:median_L_BH_div}
\end{figure*}

\begin{figure*}
  \centering
    \includegraphics[width=0.8\textwidth]{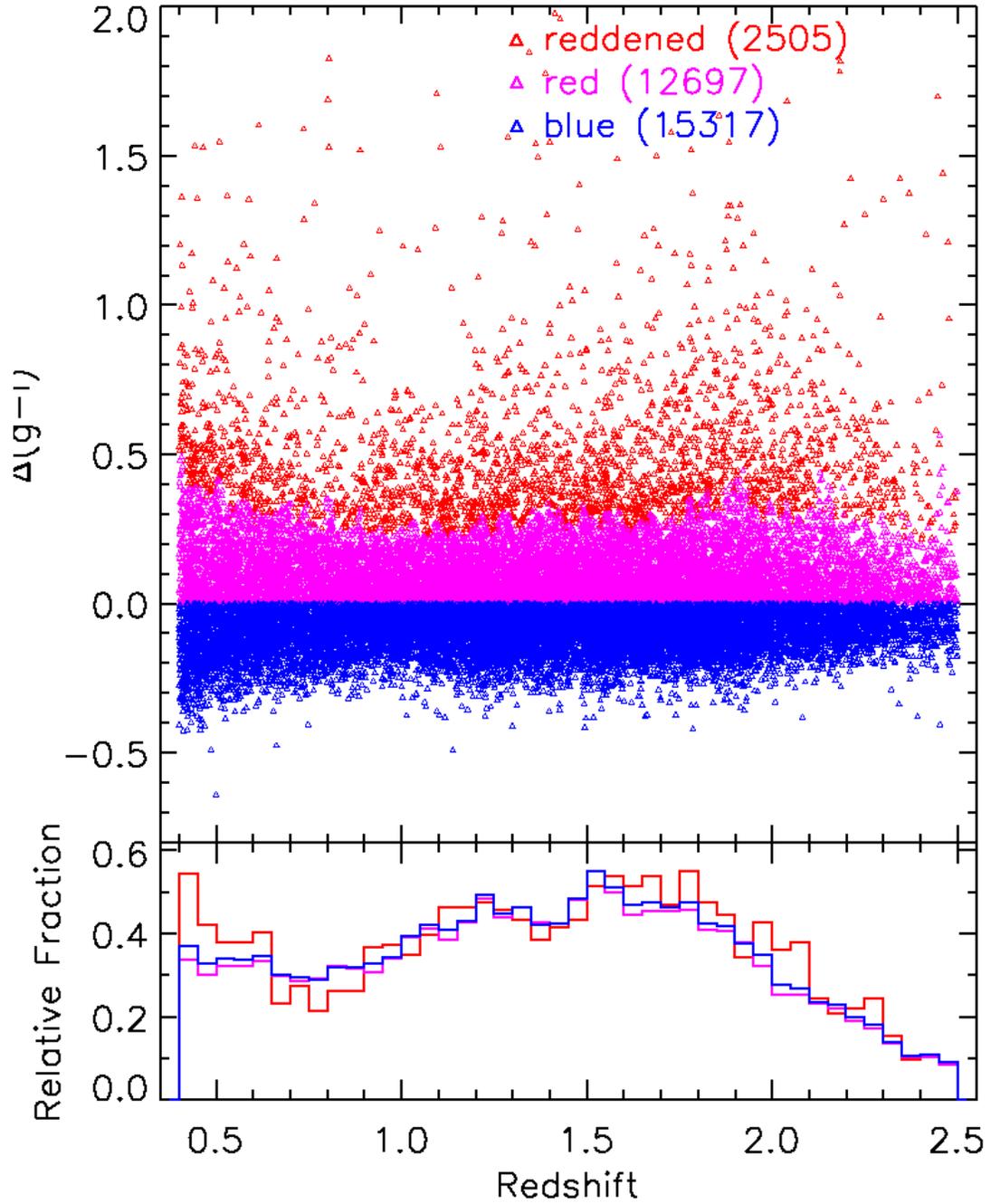}
    \caption{Sample divisions using quasar color excess (Richards et al. 2003), in the redshift
    range $0.4\le z\le 2.5$ (see \S\ref{subsec:sample}). In the bottom panel we show the normalized redshift distributions
    for each subsamples. The similarity of their redshift distributions allows a fair comparison
    of their relative clustering strength.}
    \label{fig:color_div}
\end{figure*}

\begin{figure}
  \centering
    \includegraphics[width=0.8\textwidth]{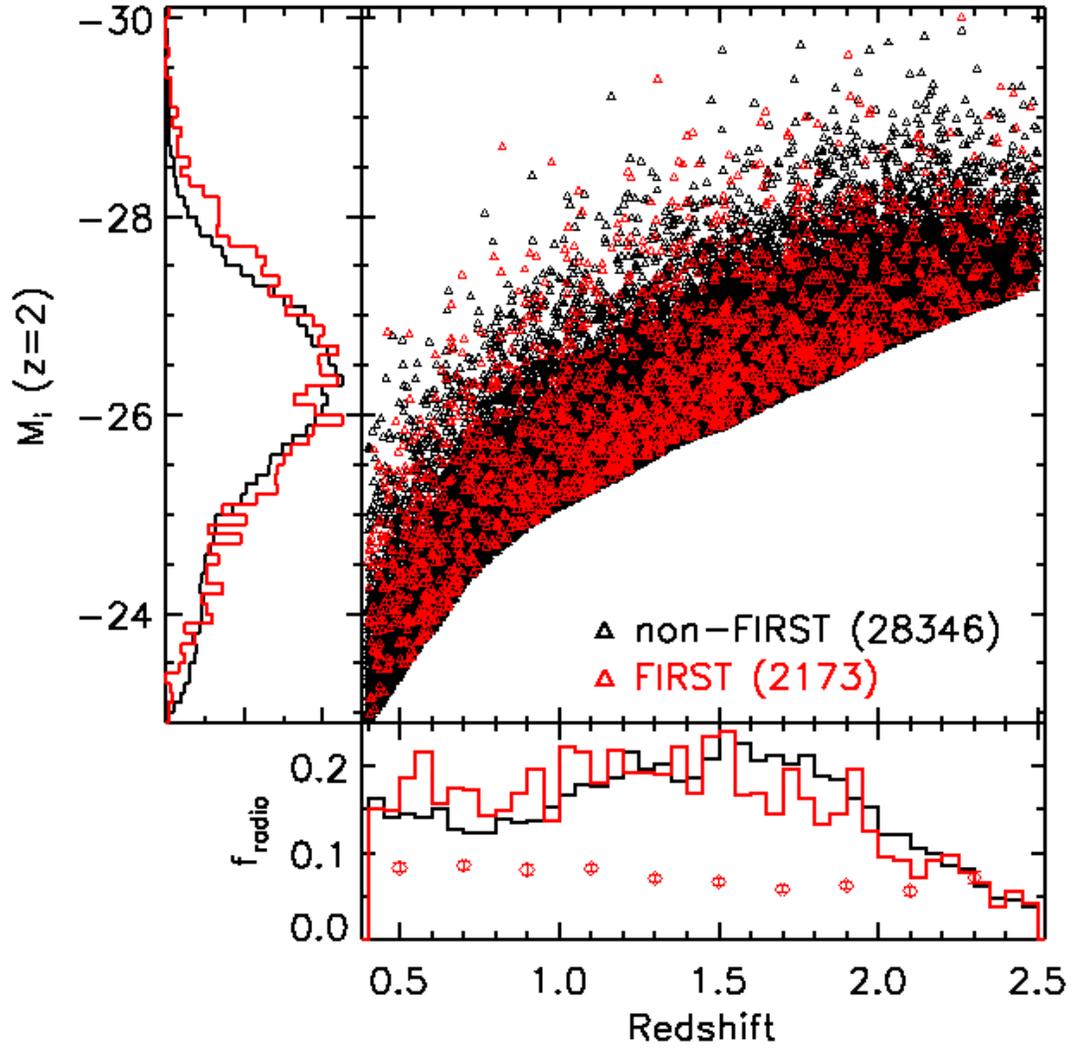}
    \caption{FIRST detected and undetected quasars. Histograms show the distributions in redshift and optical luminosity for the two
    samples. The radio fraction as a function of redshift is shown as open
    symbols in the bottom panel.}
    \label{fig:radio_div}
\end{figure}

Since all the four divisions include quasars from a wide redshift
range, it is crucial to ensure that the redshift distributions are
similar for whichever subsamples are compared. The bottom panels
in Figs. \ref{fig:median_L_BH_div}--\ref{fig:radio_div} show that
our choice of subdividing these samples yields almost identical
redshift distributions for comparing subsamples. The disadvantage
of these divisions is that they intermix quasars of different
luminosity, mass etc., at various redshifts. However, if the
dependences of clustering on these properties are monotonic and do
not evolve significantly over the redshift range $0.4\le z\le
2.5$, we still expect to see differences in the averaged sense.

\subsection{The Two-Point Correlation Function}
The simplest way to measure clustering is the redshift space
correlation function $\xi_s(s)$ where $s$ is the redshift space
separation of quasar pairs. To minimize the effects of redshift
distortions and redshift errors, the projected correlation
function $w_p(r_p)$ is often used (e.g., Davis \& Peebles 1983),
where $r_p$ is the projected separation. In our clustering
studies, both $\xi_s(s)$ and $w_p(r_p)$ will be presented. To
compute $\xi_s(s)$ and $w_p(r_p)$, we use the Landy-Szalay
estimator (Landy \& Szalay 1993). For error estimation we use
jackknife resampling as described in Shen et al. (2007a). Only the
diagonal elements in the covariance matrix are used in the
$\chi^2$ fitting because we found the covariance matrix is
generally too noisy; on the other hand, we find consistent results
with or without including off-diagonal elements in the fitting
(Shen et al. 2007a; Appendix B of Ross et al. 2008). We refer to
our previous paper (Shen et al. 2007a) and Ross et al. (2008) for
more detailed descriptions on the estimators used, error
estimations and generating random catalogs. In some cases, the
sample is too sparse to make useful auto-correlation measurements,
so we cross-correlate it with a larger quasar sample matched in
redshift to boost the clustering signal (e.g., Coil et al. 2007;
Shen et al. 2008a; Padmanabhan et al. 2008).

Although our sample represents the largest spectroscopic quasar
sample to date, we are limited by shot noise, because our sample
is much more sparse than low redshift galaxy samples (e.g., Zehavi
et al. 2005; Padmanabhan et al. 2007), type 2 AGN samples (e.g.,
Constantin \& Vogeley 2006; Li et al. 2006), or photometric quasar
samples (e.g., Myers et al. 2006, 2007a, 2007b). The volume number
density of the SDSS main quasar sample is even lower than the
number densities of the 2QZ (Croom et al. 2004) and the 2SLAQ
(Croom et al. 2008) spectroscopic samples studied by several
authors (e.g., Porciani et al. 2004; Croom et al. 2005; da
\^{A}ngela et al. 2005; da \^{A}ngela et al. 2008), because SDSS
does not reach as far down the quasar luminosity function as the
other surveys. The sparseness of the sample directly impacts our
strategy of measuring the correlation function and estimating its
error. In particular, only a handful of quasar pairs will be
present for the smallest projected $r_p$ bins when we compute
$w_p$, which not only makes the results noisy but also puts the
usage of $w_p$ into question. Binning effects may also become
important for these sparse-pair bins. For this reason, we develop
a Maximum-Likelihood method in \S\ref{subsec:ML_CF}, which we use
to fit the smallest scales.

In addition to statistical issues with our sample, there might
also be some systematics that are difficult to quantify. As
already mentioned earlier, near redshifts $z\sim 2.7$ and $z\sim
3.5$ quasars are close to the stellar locus in color-color space,
and the selection efficiency drops rapidly. The number of observed
quasars near these redshifts is reduced and pairs with similar
colors are preferentially lost, hence the correlation function
measurements suffer.
Both statistical fluctuations and systematic issues can lead to
negative correlations at certain scales. Including or excluding
these negative bins in the power-law model fitting sometimes yield
quite different results. In addition, the jackknife errors we are
using may underestimate the real errors for small number
statistics. For all these reasons, we caution that the actual
uncertainties in our clustering measurements may be larger than
our formal estimates.


\section{CLUSTERING MEASUREMENTS}
\label{sec:CF}

\subsection{Binned Correlation Functions}\label{subsec:bin_CF}

In this section we use the traditional binned correlation function
and power-law models to measure the clustering properties. We use
an alternative un-binned, maximum-likelihood approach in
\S\ref{subsec:ML_CF} for small to intermediate scales probed in
our sample.

Before we continue, we want to clarify several issues with fitting
a single power-law model of $\xi_s(s)=(s/s_0)^{-\gamma_s}$ or
$\xi(r)=(r/r_0)^{-\gamma}$ to the raw CF data points. First of
all, the underlying CF is not a perfect power-law at all scales;
even if it were, because of the statistical fluctuations of CF
data for sparse samples, the best-fit parameters of power-law
models would depend on the fitting range. Hence the best-fit
parameters should be quoted with fitting ranges. Secondly, for
most of our subsamples the data are of insufficient quality for a
flexible power-law-index fit, hence we have fixed the power-law
indices to be $\gamma_s=1.8$ and $\gamma=2$, close to the observed
values (e.g., Porciani et al. 2004; Croom et al. 2005; Shen et al.
2007a). Thirdly, occasionally some CF data points are negative due
to statistical fluctuations or unknown systematics; including
those negative points are necessary, but, because of these
outliers and the nature of power-law models, the $\chi^2$ fit will
have high model rejection probability. Hence we perform fits both
with and without negative CF data points within the fitting range.
For most of the cases the resulting correlation lengths are
consistent within $1\sigma$ errors, but there are exceptions.
Given the possibility that some of the negative points might be
caused by systematic effects (i.e., close to the redshift deserts
or at characteristic scales of preferentially losing pairs), we
believe the actual clustering strength should lie somewhere in
between these two limits.

Quasars are biased tracers of the underlying dark matter. The
linear {\em scale-independent} bias factor is normally defined as
$b^2\equiv \xi_{\rm QSO}/\xi_m$ where $\xi_m(z)$ is the matter
correlation function, extrapolated to high redshift according to
the linear growth model. We compute $\xi_m$ using a linear CDM
power spectrum with transfer function from Eisenstein \& Hu
(1999), and estimate $b$ using the integrated correlation function
to marginalize nonlinear effects and scale-dependent
bias\footnote{We find that the bias calculated in this way is only
slightly larger than the one estimated at $r=20\ h^{-1}$Mpc by
$\sim 5\%$, below the uncertainty from the clustering measurement
itself ($\gtrsim 10\%$; see Table 1).}
\begin{equation}
\xi_{20}=\frac{3}{r_{\rm max}^3}\int_{r_{\rm min}}^{r_{\rm
max}}\xi(r)r^2dr\ ,
\end{equation}
where we choose $r_{\rm min}=5\ h^{-1}{\rm Mpc}$ and $r_{\rm
max}=20\ h^{-1}{\rm Mpc}$ (note these choices are slightly
different from Ross et al. 2008). We use the best-fit power-law
models to compute the integrated $\xi_{20}$ for quasars, and
estimate the matter correlation function at the median redshift of
the quasar sample. For cross-correlation functions with power-law
models and fixed power-law index $\gamma=2$, we simply have
\begin{equation}\label{eqn:cross_CF}
r_{0,2}=r_{{\rm 0,cross}}^2/r_{0,1}\ ,\ b_{2}=b_{\rm cross}^2/b_1\
,
\end{equation}
where subscript ``cross'' refers to the cross-correlation and $1$,
$2$ refer to auto-correlations of sample 1 and 2. Note that with
equation (\ref{eqn:cross_CF}) we have implicitly assumed that both
samples 1 and 2 are perfectly correlated with each other, i.e.,
they both trace the underlying DM distribution exactly.

\begin{deluxetable*}{lcccccccccc}
\tabletypesize{\footnotesize}  
\tablecolumns{11} \tablewidth{1.0\textwidth}
\tablecaption{Measurements of the two-point correlation function
\label{table:CF_result}} \tablehead{Sample & $N_{\rm QSO}$ &
$\bar{z}$ & $\log_{10}\bar{L}_{\rm bol}$ &$[s_1,s_2]$ &
$s_0|_{\gamma_s=1.8}$
 & $[r_{p,1},r_{p,2}]$  &
$r_0|_{\gamma=2}$ & $b$ & ML $[R_1,R_2]$ & $r_{\rm
0,ML}|_{\gamma_{\rm ML}=2}$ \\
  & & & $({\rm erg\,s^{-1}})$ & ($h^{-1}$Mpc) & ($h^{-1}$Mpc) &
  ($h^{-1}$Mpc) & ($h^{-1}$Mpc) & & ($h^{-1}$Mpc) & ($h^{-1}$Mpc)
 }  \startdata
\textbf{All luminosities} & & & & & & & & & & \\
$0.1<z<0.8$ & 7902   &0.50 & 45.57 & $[2,40]$   & $5.07\pm0.40$    & $[3,40]$   & $6.14\pm0.78$     &$1.32\pm0.17$ & $[2,15]$  & $6.94_{-0.73}^{+0.62}$\\
                       &    &     &  &            &                   &            &                   &              & $[5,20]$  & $7.25_{-0.90}^{+0.78}$\\
$0.8<z<1.4$ & 9975   &1.13 & 46.27  & $[2,150]$  & $6.21\pm0.59$    & $[3,80]$   & $8.16\pm0.79$     &$2.31\pm0.22$ & $[2,15]$  & $6.53_{-1.14}^{+1.01}$\\
                       &     &     &  &           & $(6.01\pm0.60)$  &            & $(7.76\pm0.80)$   &$2.20\pm0.23$ & $[5,20]$  & $8.18_{-1.16}^{+1.03}$\\
$1.4<z<2.0$ & 11304   &1.68 & 46.60  & $[2,150]$  & $7.35\pm0.55$    & $[3,80]$   & $8.51\pm0.76$     &$2.96\pm0.26$ & $[2,15]$  & $7.82_{-0.97}^{+0.90}$\\
                       &    &    &    &        & $(6.94\pm0.56)$  &            & $(7.16\pm0.77)$   &$2.49\pm0.27$ & $[5,20]$  & $9.37_{-0.98}^{+0.93}$\\
$2.0<z<2.5$ & 3828   &2.18 & 46.86  & $[2,150]$  & $8.36\pm1.38$    & $[3,80]$   & $11.47\pm1.71$    &$4.69\pm0.70$ & $[2,15]$  & $6.05_{-3.31}^{+2.18}$\\
                       &    &    &    &        & $(7.40\pm1.42)$  &            & $(8.04\pm1.94)$   &$3.29\pm0.79$ & $[5,20]$  & $5.41_{-3.21}^{+2.71}$ \\
$2.9<z<3.5$ & 2693   &3.17 & 46.73  & $[2,150]$  & $10.36\pm1.82$   & $[16,150]$ & $14.58\pm2.70$    &$7.76\pm1.44$ & $[2,15]$  & $6.23_{-3.80}^{+3.26}$\\
                       &    &    &    &        & $(8.82\pm1.86)$  &            & $(13.34\pm2.78)$  &$7.10\pm1.48$ & $[5,20]$  & $0.03_{-0.03}^{+5.82}$\\
                       &    &    &    &        &                  &            &                   &     & $[20,40]$ & $19.60_{-4.76}^{+3.50}$ \\
$3.5<z<5.0$ & 1788   &3.84 & 47.07  &$[16,150]$ & $22.10\pm3.44$   & $[16,150]$ & $21.04\pm3.39$    &$12.96\pm2.09$& $[2,15]$  & $17.54_{-5.44}^{+4.58}$\\
                       &    &    &    &        & $(17.05\pm3.61)$ &            & $(16.00\pm 3.68)$ &$9.85\pm2.27$ & $[5,20]$  & $21.04_{-5.03}^{+4.45}$\\
\hline\\
$2.9<z<3.5$  & & & & & & & & & &\\
$-27.3<M_i<-25.7$  & 1262       &3.17 & 46.67  &  &                  &            &                   & & $[2,15]$  & $0.00_{-0.00}^{+8.33}$ \\
                   &        &     &   &        &                  &            &                   & & $[5,20]$  & $0.02_{-0.02}^{+7.16}$\\
                   &        &    &    &        &                  &            &                   & & $[20,40]$ & $21.92_{-9.54}^{+6.46}$\\
$-29.7<M_i<-27.3$  & 1416       &3.17 & 46.91 &      &                  &            &                   & & $[2,15]$  & $11.04_{-6.29}^{+5.05}$ \\
                   &        &    &    &        &                  &            &                   & & $[5,20]$  & $13.31_{-6.90}^{+5.14}$\\
                   &        &    &    &        &                  &            &                   & & $[20,40]$ & $12.60_{-7.86}^{+7.49}$\\
\hline\\
$0.4<z<2.5$        &        &    & & & & & & & &\\
bright  & 15291     &1.40 & 46.56     & $[2,115]$  & $7.71\pm0.67$    & $[3,115]$  & $7.19\pm1.06$     &$2.26\pm0.33$ &           &\\
        &                   &  &    &            &$(6.38\pm0.66)$   &            & $(6.42\pm1.10)$   &$2.02\pm0.35$ &           &\\
faint   & 15228      &1.40 & 46.31     & $[2,115]$  & $8.11\pm0.64$    & $[3,115]$  & $6.86\pm0.75$     &$2.16\pm0.24$ &           &\\
        &                   &    &    &            & $(7.60\pm0.64)$  &            &              &              &           &\\
$10\%$ most luminous-cross & 3050 &1.40 & 46.84   & $[2,115]$  & $7.78\pm0.82$    & $[3,115]$  & $9.53\pm0.80$     & $3.00\pm0.25$ &           &\\
                          &      &     &     &            &                  &            & $(8.72\pm0.82)$   & $2.75\pm0.26$  &           &\\
$90\%$ least luminous & 27469 &1.40    & 46.39   & $[2,115]$  &  $6.39\pm0.37$      & $[3,115]$  & $7.06\pm0.46$     & $2.22\pm0.14$ &           &\\
$10\%$ most luminous-auto & 3050 &1.40 & 46.84    &         &                    &             & $12.86\pm2.32$    & $4.05\pm0.73$     &           &\\
                         &   &     &    &        &                 &             & $(10.77\pm2.14)$    & $3.39\pm0.67$     &           &\\
high-mass  & 13414        &1.35 & 46.51   &      $[2,115]$  & $8.25\pm0.55$    & $[3,115]$  & $8.44\pm0.82$     &$2.61\pm0.25$ &           &\\
           &              &  &    &            & $(7.36\pm0.56)$  &            &$(7.78\pm0.85)$    &$2.40\pm0.26$ &           &\\
low-mass   & 13443        &1.35 & 46.37 &     $[2,115]$  & $5.70\pm0.68$    & $[3,115]$  & $8.00\pm1.00$     &$2.47\pm0.31$ &           &\\
           &              &  &    &            & $(5.56\pm0.68)$  &            & $(7.29\pm1.01)$   &$2.25\pm0.31$ &           &\\
$10\%$ most massive-cross & 2698 & 1.35 &46.64  & $[2,115]$  & $8.62\pm0.87$    & $[3,115]$  & $8.73\pm0.88$   & $2.70\pm0.27$  &           &\\
                         &      &     &  &          & $(8.23\pm0.88)$  &            &                 &                &           &\\
$90\%$ least massive & 24159 &1.35 & 46.42  &$[2,115]$  & $6.53\pm0.44$     & $[3,115]$  & $7.10\pm0.57$  & $2.19\pm0.18$ &           &\\
$10\%$ most massive-auto & 2698 &1.35   & 46.64  &             &                 &              & $10.73\pm2.33$   & $3.32\pm0.72$       &           &\\
reddened-cross & 2505  & 1.40 & 46.31  &$[2,115]$  & $6.08\pm0.94$    & $[3,115]$  & $8.10\pm1.20$     &$2.55\pm0.38$ &           &\\
               &         &   &    &            & $(5.31\pm0.98)$  &            & $(6.10\pm1.36)$   &$1.92\pm0.43$ &           &\\
red  & 12697         &1.40 & 46.41  &  $[2,115]$  & $7.71\pm0.80$    & $[3,115]$  & $9.17\pm0.93$     &$2.89\pm0.29$ &           &\\
     &                   &   &    &            &         &            & $(8.42\pm0.95)$   &$2.65\pm0.30$ &           &\\
blue & 15317        &1.40 &46.43   &  $[2,115]$  & $7.68\pm0.60$    & $[3,115]$  & $7.81\pm0.71$     &$2.46\pm0.22$ &           &\\
FIRST-cross & 2173  &1.30 &46.38   &  $[2,115]$  & $8.91\pm0.87$    & $[3,115]$  & $9.61\pm0.85$     &$2.91\pm0.26$ &           &\\
            &            &   &    &            & $(8.68\pm0.88)$  &            & $(9.18\pm0.86)$   &$2.78\pm0.26$ &           &\\
non-FIRST   & 28346   &1.40 &46.42 &  $[2,115]$  & $6.91\pm0.36$    & $[3,115]$  & $7.12\pm0.50$     &$2.24\pm0.16$ &           &\\
            &  &     &  &          & $(6.64\pm0.36)$  &            &                   &              &           &\\
FIRST-auto  & 2173  &1.30 & 46.38  &         &                  &            &  $12.97\pm2.47$   & $3.93\pm0.75$            &           &\\
            &  &     &   &         &                  &            &  $(11.84\pm2.37)$   & $3.59\pm0.72$           &           &\\
\enddata
\tablecomments{Tabulated values of the fitted correlation
functions for various samples. All length scales are in units of
$h^{-1}$Mpc. The fitting ranges are shown in ``$[\ ]$'', and in
``$(\ )$'' we show the fitting results including negative data
points (see the discussion in the text). The $1\sigma$ error
$\delta b$ in $b$ is just $\delta_b=b(\delta r_0/r_0)$ for the
fixed power-law $\gamma=2$ models and neglecting the theoretical
uncertainty in the matter correlation function $\xi_m(r,z)$.}
\end{deluxetable*}

\subsubsection{Redshift Evolution}
In the companion paper (Ross et al. 2008) we study the redshift
evolution of quasar clustering in detail. Here we only provide
brief discussion on this topic.

There have been several studies of the redshift evolution of
quasar clustering, based on large survey samples such as 2QZ,
SDSS, and 2SLAQ (e.g., Croom et al. 2005; Myers et al. 2006,
2007a; Porciani \& Norberg 2006; Shen et al. 2007a; da \^{A}ngela
et al. 2008; Ross et al. 2008). While the redshift evolution at
$z\lesssim 2.2$ is still controversial to some extent (cf.
Porciani \& Norberg 2006; da \^{A}ngela et al. 2008; Ross et al.
2008) owing to the uncertainties in the clustering measurements
for individual redshift bins, high redshift ($z\gtrsim 3$) SDSS
quasars have much stronger clustering than their low redshift
($z\lesssim 2.2$) counterparts (Shen et al. 2007a), reflecting the
large bias of their host halos.

We have performed independent clustering measurements using our
redshift bins indicated in the left panel of Fig.
\ref{fig:z_L_BH_div}, but with all the quasars in each redshift
bin, i.e., no luminosity cut. The best-fit values of correlation
length and linear bias are tabulated in Table
\ref{table:CF_result}. Our results are in good agreement with
those of Ross et al. (2008). We see a slight trend of increasing
clustering with redshift at $z<2.5$, but the uncertainties are
large in each redshift bin.


\subsubsection{Luminosity Dependence}\label{subsec:L_dep}
At each redshift our sample spans a range in luminosity (typically
$3$ magnitudes). Hence the next question is whether or not quasar
clustering depends on luminosity. On theoretical grounds, if
quasar luminosity perfectly correlates with host halo mass, we
expect that higher-luminosity quasars live in more massive halos
and therefore are more strongly clustered. Weak dependence of
quasar clustering on luminosity would imply that a quasar's
luminosity is not a good tracer of its halo mass -- there is
substantial scatter in $L-M_{\rm BH}$ and/or $M_{\rm BH}-M_{\rm
halo}$ correlations (e.g., Lidz et al. 2006). Observationally an
important caveat is that the errors on the clustering measurements
must be small enough that any difference caused by varying host
halo mass is discernible. No strong luminosity dependence has yet
been reported (e.g., Adelberger \& Steidel 2005; Porciani \&
Norberg 2006; Myers et al. 2007a; da \^{A}ngela et al. 2008); but
most of the studies are for $z\lesssim 2.2$ and the error bars are
large.

\begin{figure*}
  \centering
    \includegraphics[width=0.45\textwidth]{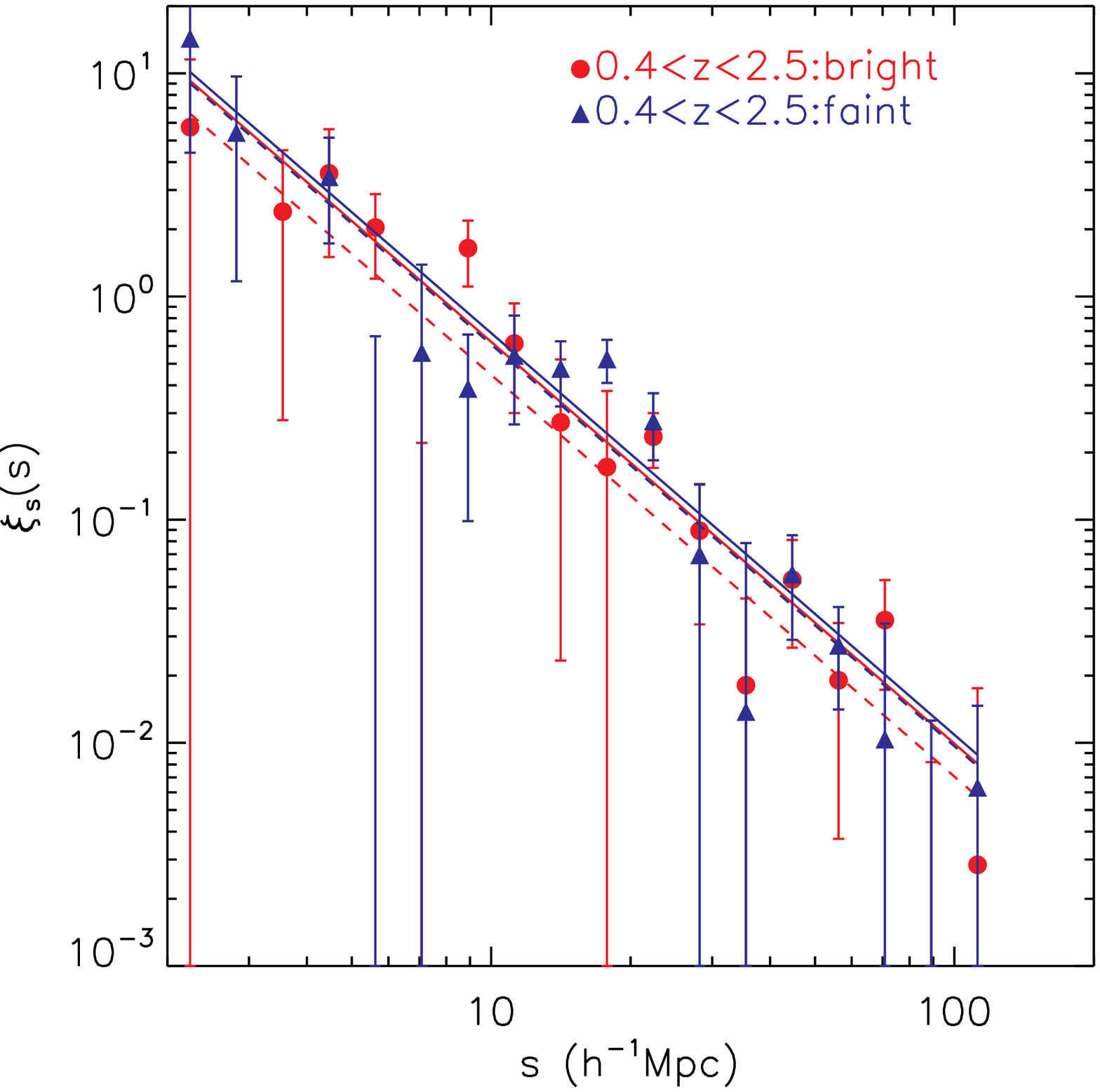}
    \includegraphics[width=0.45\textwidth]{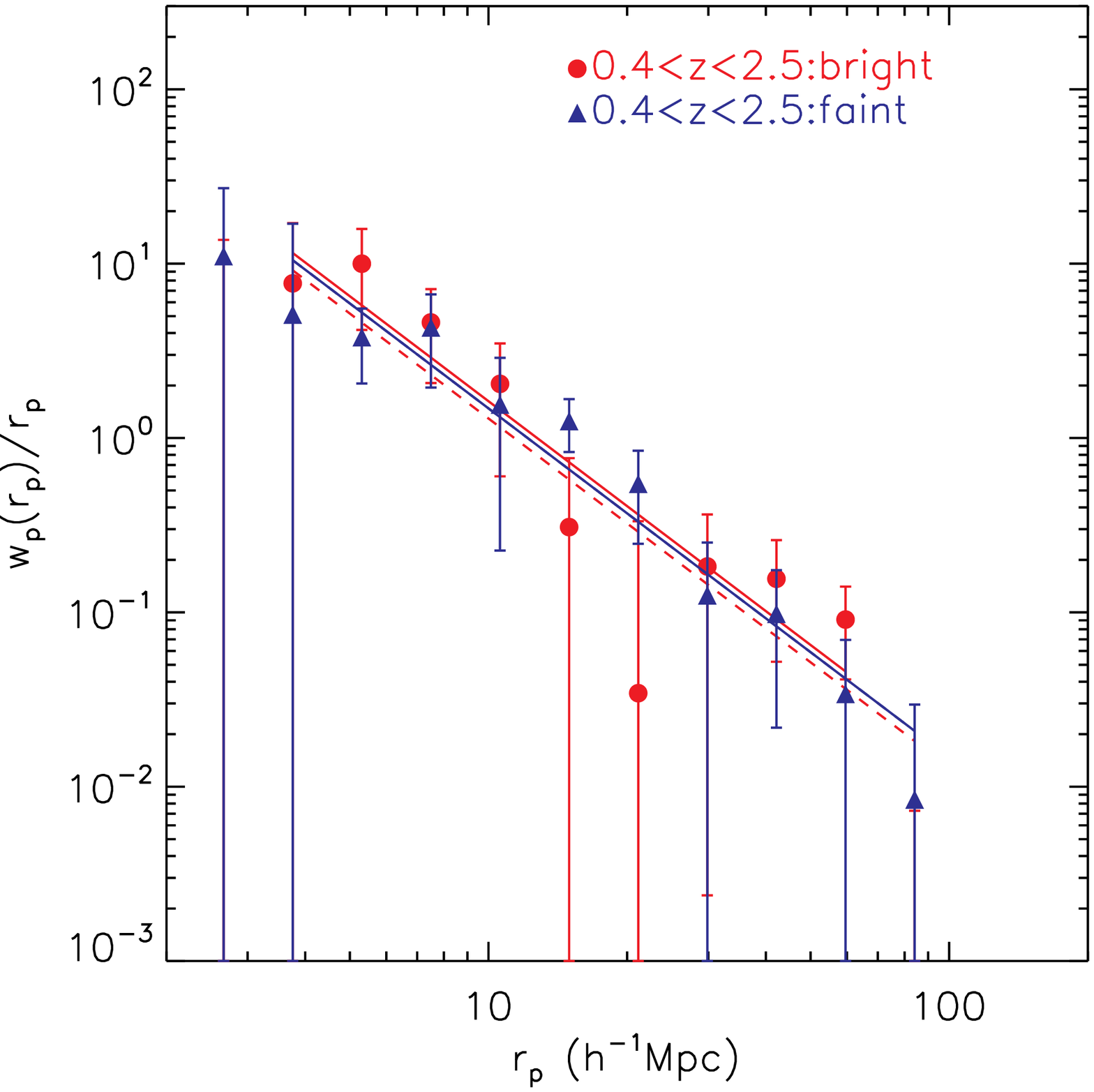}
    \includegraphics[width=0.45\textwidth]{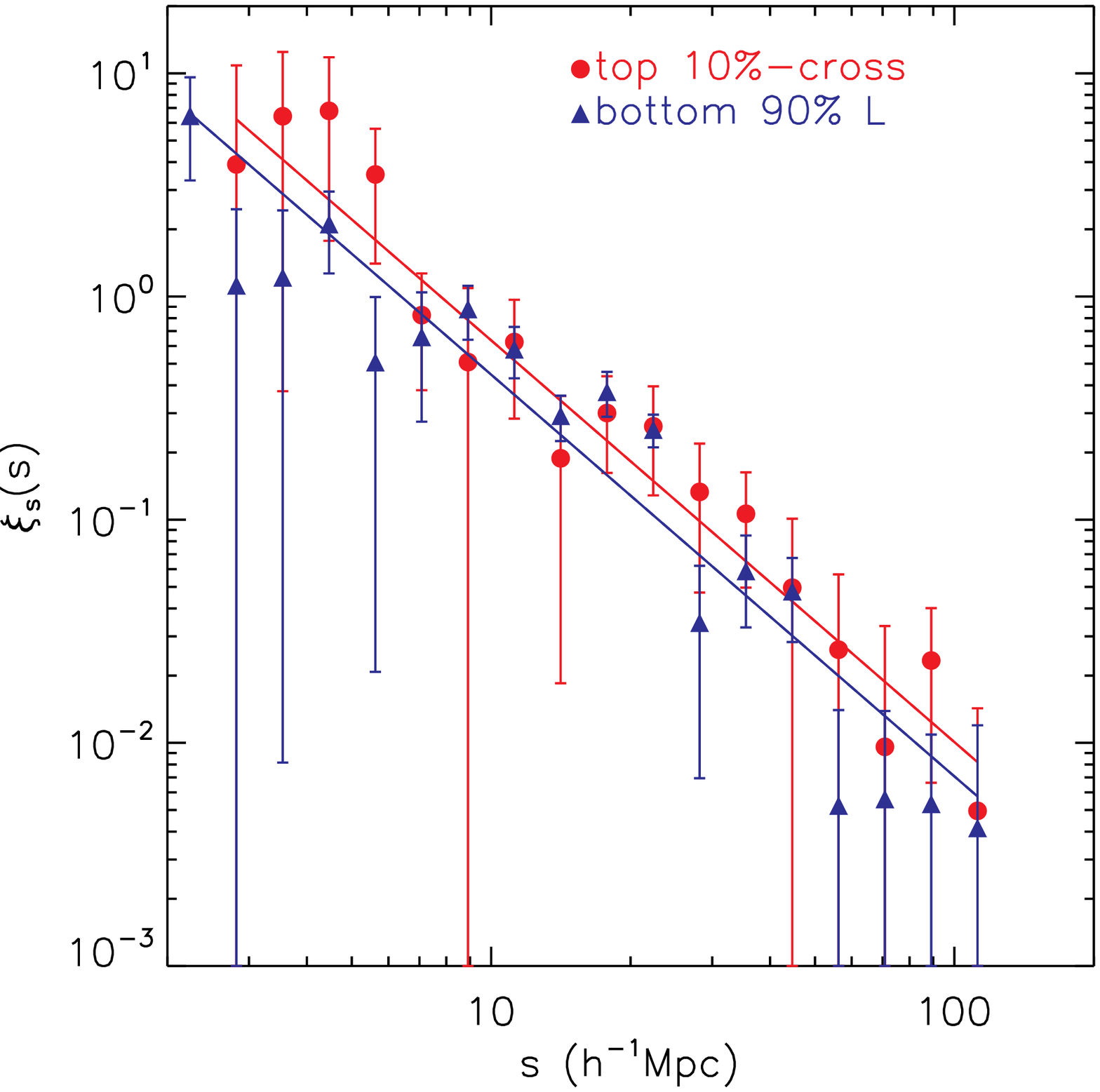}
    \includegraphics[width=0.45\textwidth]{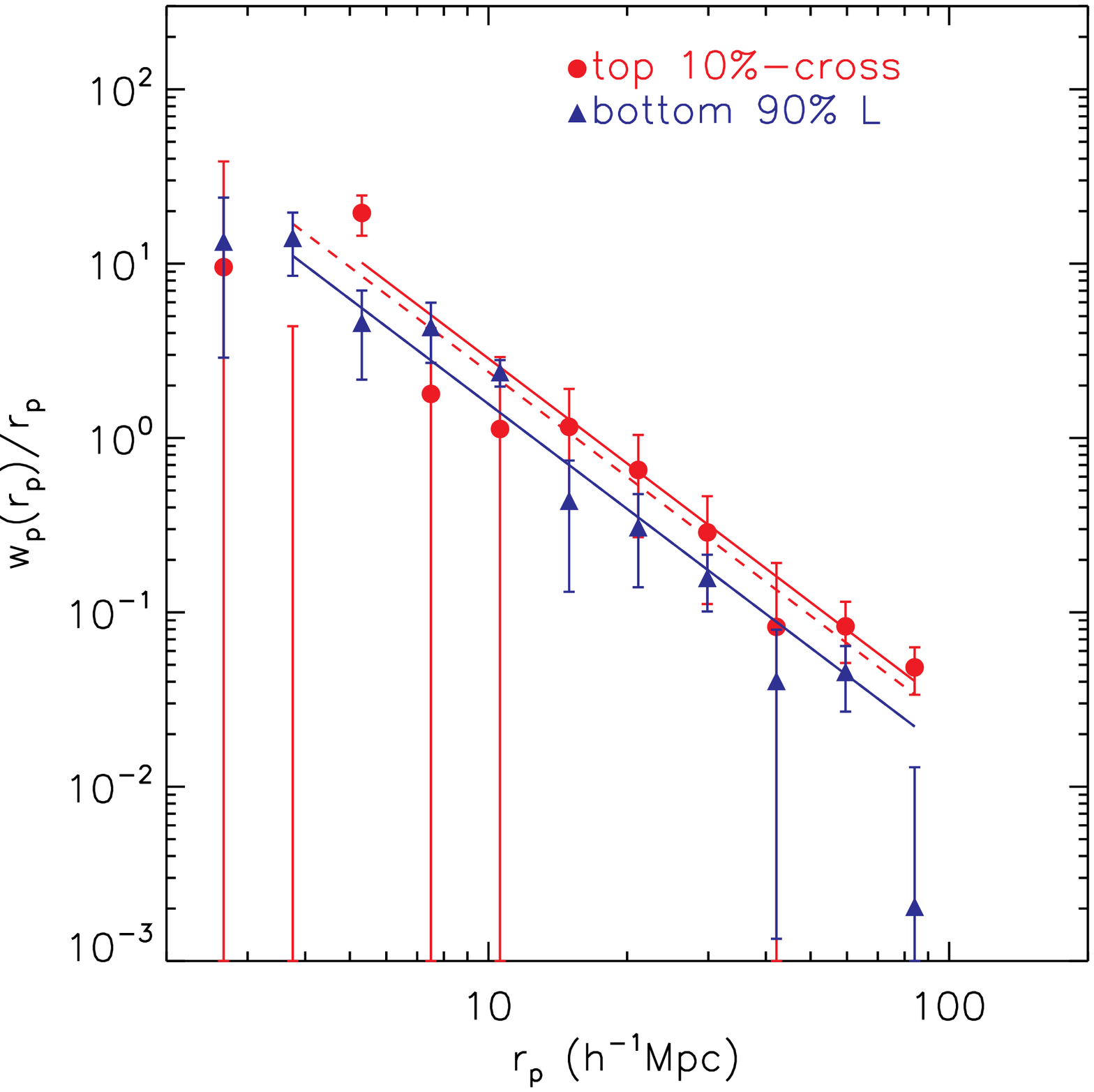}
    \caption{Luminosity dependence of quasar clustering, using the bright/faint
    samples (top panels) or the top $10\%$ luminosity divided
    samples (bottom panels), with $0.4\le z\le 2.5$ as defined in
    \S\ref{subsec:sample}. For the top $10\%$ most luminous quasars we have cross-correlated them with
    the remaining $90\%$ quasars. Solid and dashed lines are the best-fit
    power-law models excluding and including negative data bins
    in the fits, respectively (see text for details).
    {\em Left:} redshift space correlation
    function. {\em right:} projected correlation function. No appreciable difference in the clustering
    strength is observed for the median luminosity division. But the most luminous quasars show appreciably
    stronger clustering than the rest of the quasars.}
    \label{fig:L_dep}
\end{figure*}

We attack this problem by first examining the $0.4\le z\le 2.5$
samples divided by the median luminosity at each redshift
(\S\ref{subsec:sample} and the top-left panel of Fig.
\ref{fig:median_L_BH_div}), which offers the best signal-to-noise
ratio, although it mixes objects at different redshifts and
different luminosities. The correlation function for the {\em
bright} and {\em faint} quasar samples are shown in the upper two
panels in Fig. \ref{fig:L_dep}. The projected correlation
functions are consistent with one another given the errors.

We then compare the results for the $10\%$ most luminous quasars
and the remainder, as shown in the bottom panels in Fig.
\ref{fig:L_dep}, where we use cross-correlation for the $10\%$
most luminous quasars with the remaining $90\%$ less luminous
quasars. We detect stronger clustering strength for the most
luminous quasars at the $\sim 2\sigma$ level (e.g., see Table
\ref{table:CF_result}). This is somewhat expected since the halo
bias factor increases rapidly with mass. If these brightest
quasars live in the rarest and most massive halos, their
clustering will be appreciably stronger than the fainter quasars,
an effect that is detectable even with our current sample size.

\begin{figure*}
  \centering
    \includegraphics[width=0.6\textwidth,angle=90]{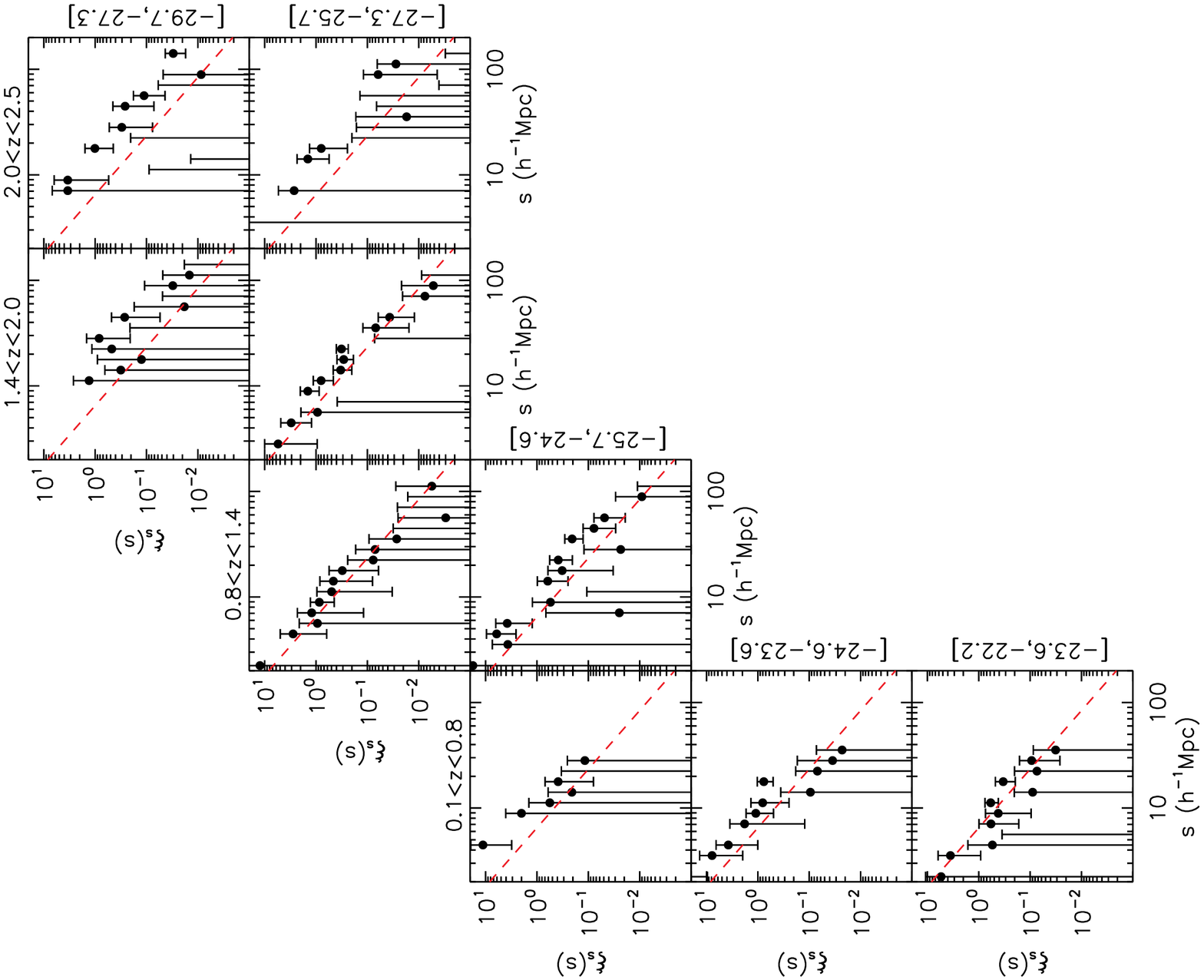}
    \includegraphics[width=0.6\textwidth,angle=90]{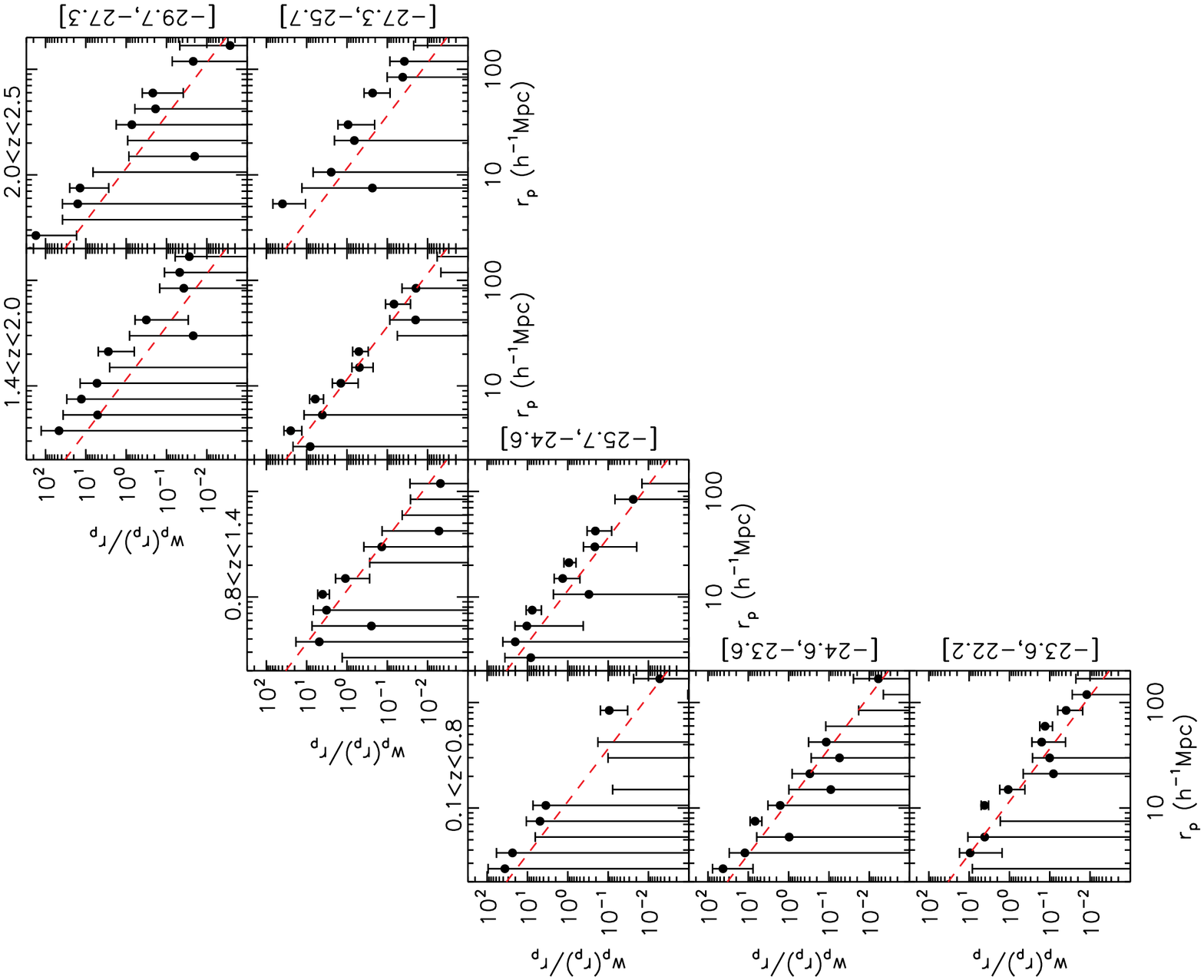}
    \caption{Correlation functions in each redshift-luminosity bin as defined
    in the left panel of Fig. \ref{fig:z_L_BH_div}. The redshift
    ranges are marked on the top of each column and
    the luminosity [$M_i(z=2)$] ranges
    are marked on the right of each row. {\em Left:} redshift space
    correlation function. {\em Right:} projected correlation function. In each
    panel, the dashed line is a power law model of $\xi_s(s)=(s/6.5\ h^{-1}{\rm Mpc})^{-1.8}$ or
    $\xi(r)=(r/6.5\ h^{-1}{\rm Mpc})^{-2}$, drawn
    to guide the eye. No appreciable difference is seen in each
    luminosity bin at fixed redshift, due to large error bars in
    the correlation functions caused by small sample sizes.}
    \label{fig:Lz_fine}
\end{figure*}

Finally we compare the clustering in each luminosity-redshift bin
as indicated in the left panel of Fig. \ref{fig:z_L_BH_div}, since
the luminosity dependence of clustering may evolve with redshift.
The results are shown in Fig. \ref{fig:Lz_fine}. Given the large
uncertainties, no appreciable luminosity dependence is observed in
any of the redshift bins. The comparison of different luminosity
samples becomes increasingly difficult at high redshift, and hence
we defer the discussion of the luminosity dependence of clustering
at $z\ge 2.9$ to \S\ref{subsec:ML_CF}.

For the overall quasar population, one can also compare the
clustering amplitudes between our sample and the 2QZ/2SLAQ samples
(e.g, Porciani et al. 2004; Croom et al. 2005; Porciani \& Norberg
2006; da \^{A}ngela et al. 2008) or the SDSS photometric quasar
samples (e.g., Myers et al. 2006, 2007a). For the same redshift
ranges at $z\lesssim 2.2$, all samples have comparable clustering
amplitude within the error bars. The average luminosity is the
highest for our SDSS spectroscopic quasar sample and the lowest
for the 2SLAQ sample (differing by a factor of {\bf $\sim 5$}),
which also indicates weak luminosity dependence of quasar
clustering at $z\lesssim 2.2$ for all but the highest
luminosities. We discuss the implications of this weak luminosity
dependence at $z< 2.5$ in \S\ref{sec:diss}.

\subsubsection{Virial Mass Dependence}\label{subsec:BH_vir_dep}
We have divided the sample in virial BH masses using the
measurements from Shen et al. (2008b). Quasars have a range of
Eddington ratios at fixed BH mass (e.g., Kollmeier et al. 2006;
Shen et al. 2008b), thus the instantaneous quasar luminosity is
not a good tracer of the BH mass. If BH mass is a superior tracer
of its host halo mass, we might expect a BH mass dependence of
quasar clustering. We test this possibility using virial BH masses
measured from single-epoch quasar spectra. However, we note that
virial BH masses are {\em not} true BH masses (see the extensive
discussion in Shen et al. 2008b). The scatter between virial and
true BH masses may also dilute any potential clustering
dependence, which we will return to in
\S\ref{subsec:BH_vir_vs_Mhalo}.

\begin{figure*}
  \centering
    \includegraphics[width=0.45\textwidth]{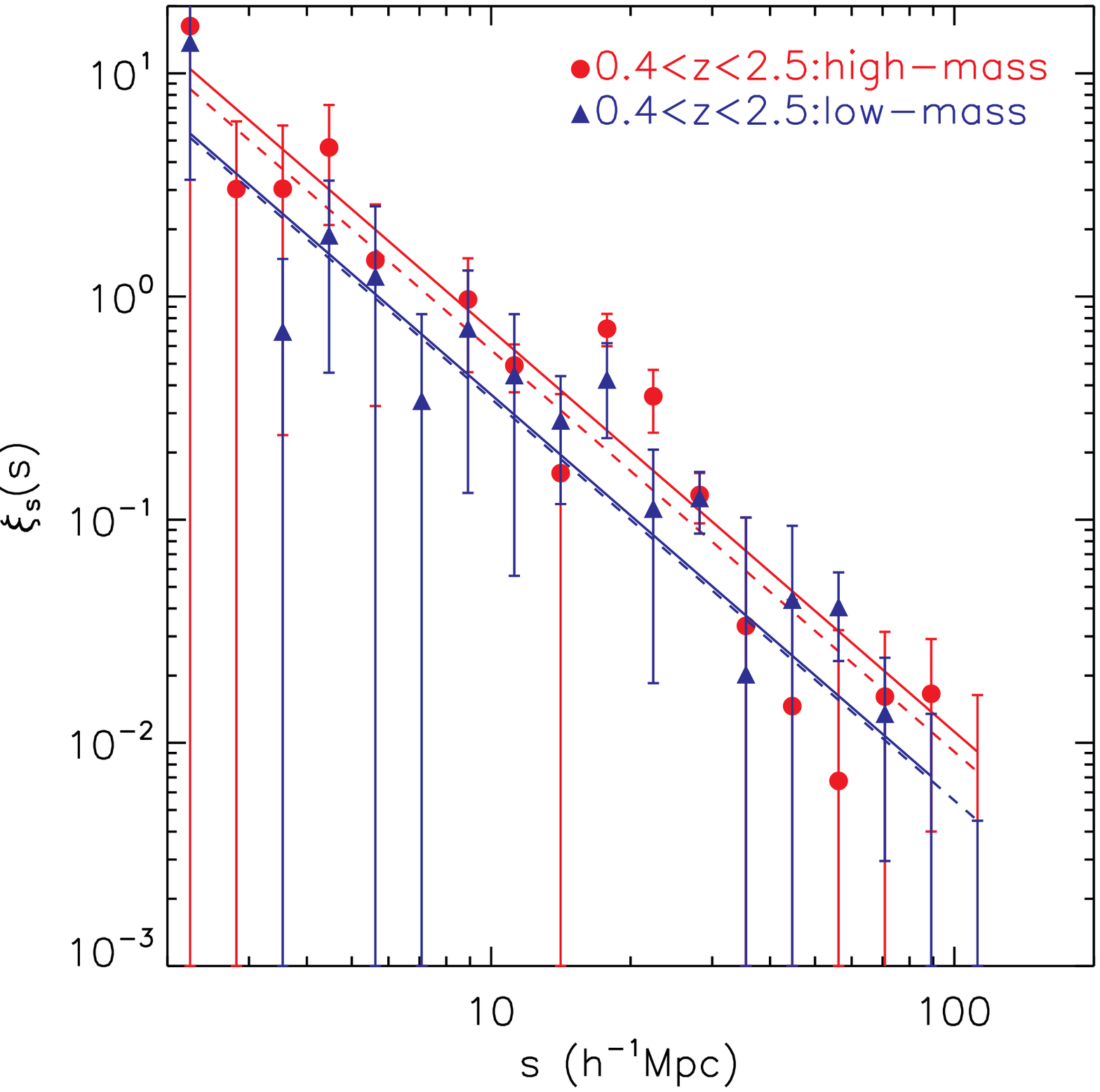}
    \includegraphics[width=0.45\textwidth]{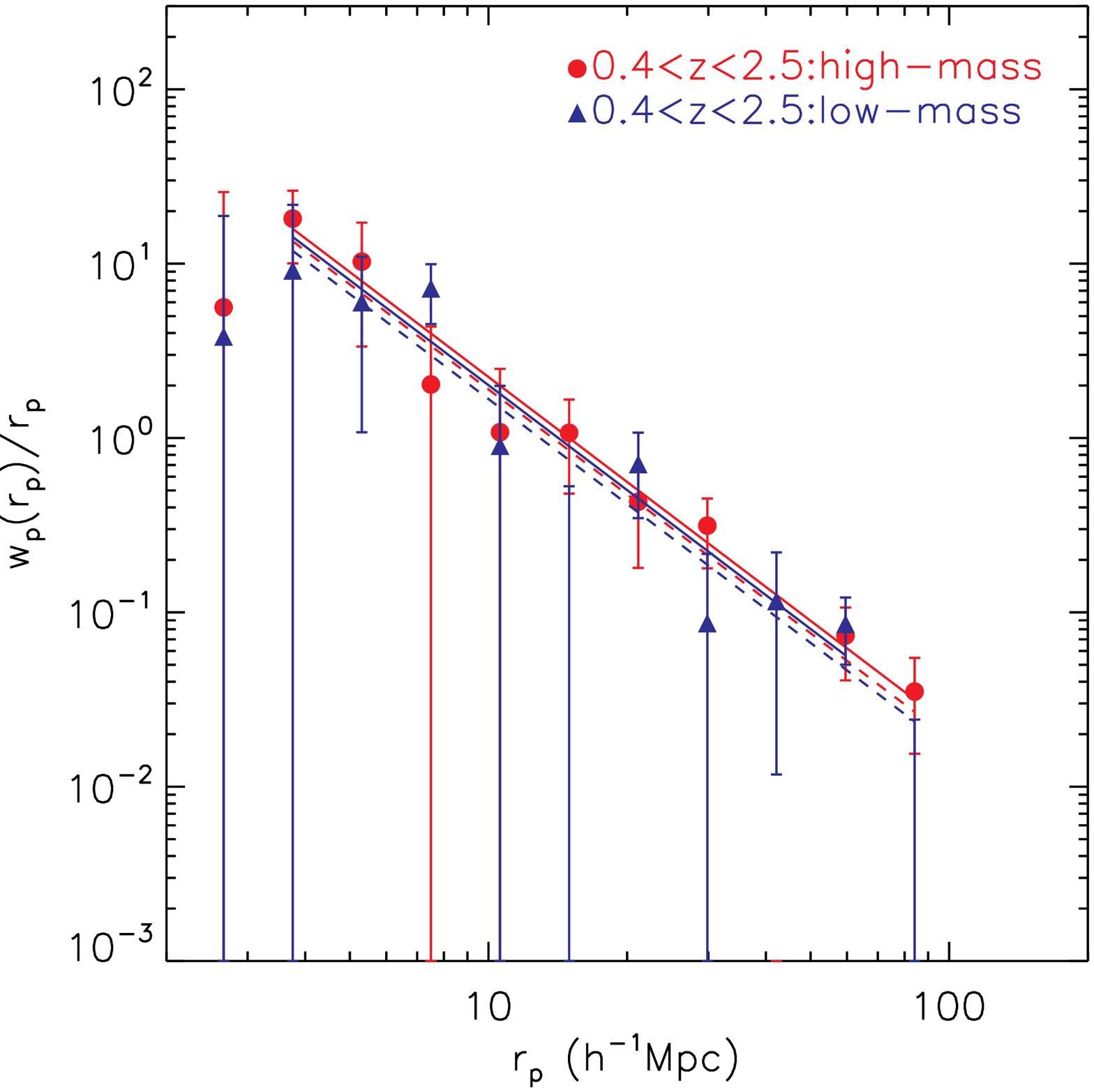}
    \includegraphics[width=0.45\textwidth]{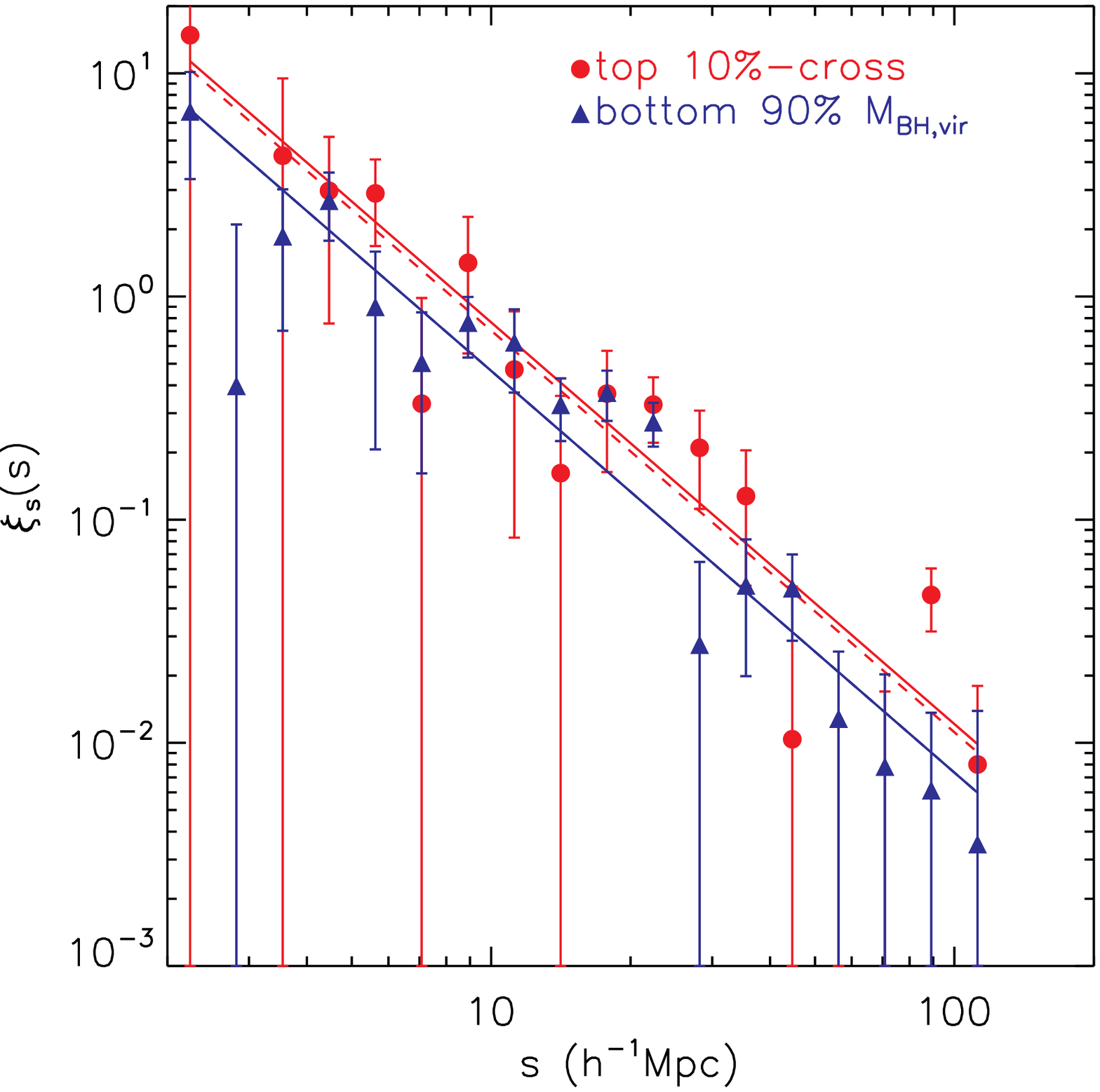}
    \includegraphics[width=0.45\textwidth]{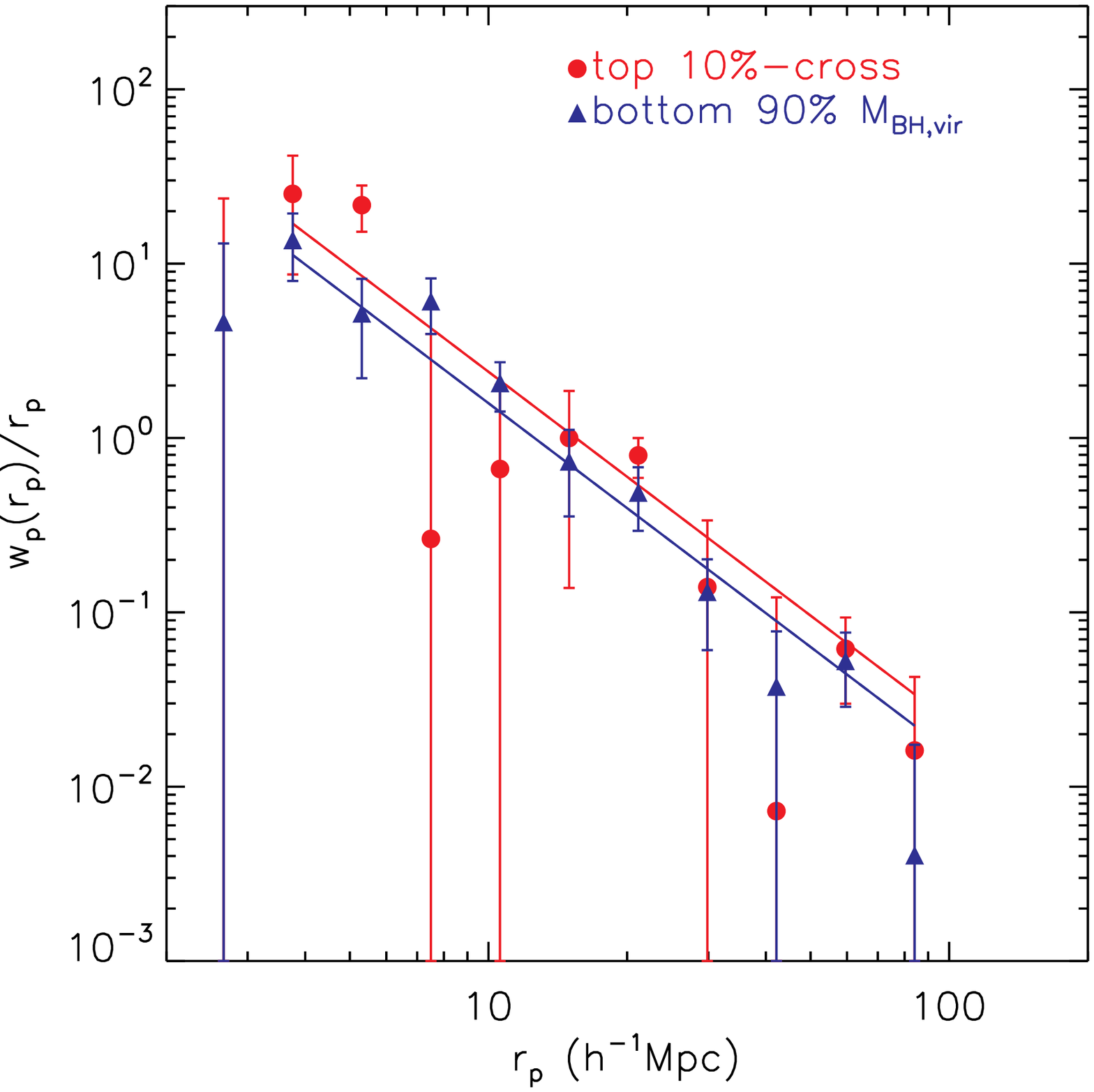}
    \caption{Virial mass dependence of quasar clustering, using the {\em high}/{\em low-mass}
    samples (upper panels) or the top $10\%$ virial mass divided samples (lower panels) as defined in \S\ref{subsec:sample}.
    Solid and dashed lines are the best-fit
    power-law models excluding and including negative data bins
    in the fits, respectively.
    {\em Left:} redshift space correlation.
    {\em Right:} projected correlation function. The {\em high/low-mass} samples are consistent with no difference
    in clustering strength given the measurement errors. However, the most massive quasars are more strongly clustered
    than are the rest of the quasars.}
    \label{fig:BH_dep}
\end{figure*}

As in the luminosity dependence case, we show the comparison of
our $0.4\le z\le 2.5$ {\em high mass} and {\em low mass} samples
in Fig. \ref{fig:BH_dep}. While there is some indication that the
{\em high mass} sample has a stronger clustering amplitude in
redshift space, it is completely consistent with no difference in
the projected correlation function (see Table
\ref{table:CF_result} for the best-fit $r_0$ values). However,
when the sample is divided into the top $10\%$ massive quasars and
the remainder, the clustering is stronger for the quasars with
highest virial BH masses, as shown in the bottom panels of Fig.
\ref{fig:BH_dep}. The difference in the clustering strength for
the cross-correlation of the $10\%$ most massive quasars and the
auto-correlation of the remaining $90\%$ is significant at the
$\sim 2\sigma$ level (see Table \ref{table:CF_result}).

\begin{figure*}
  \centering
    \includegraphics[width=0.7\textwidth]{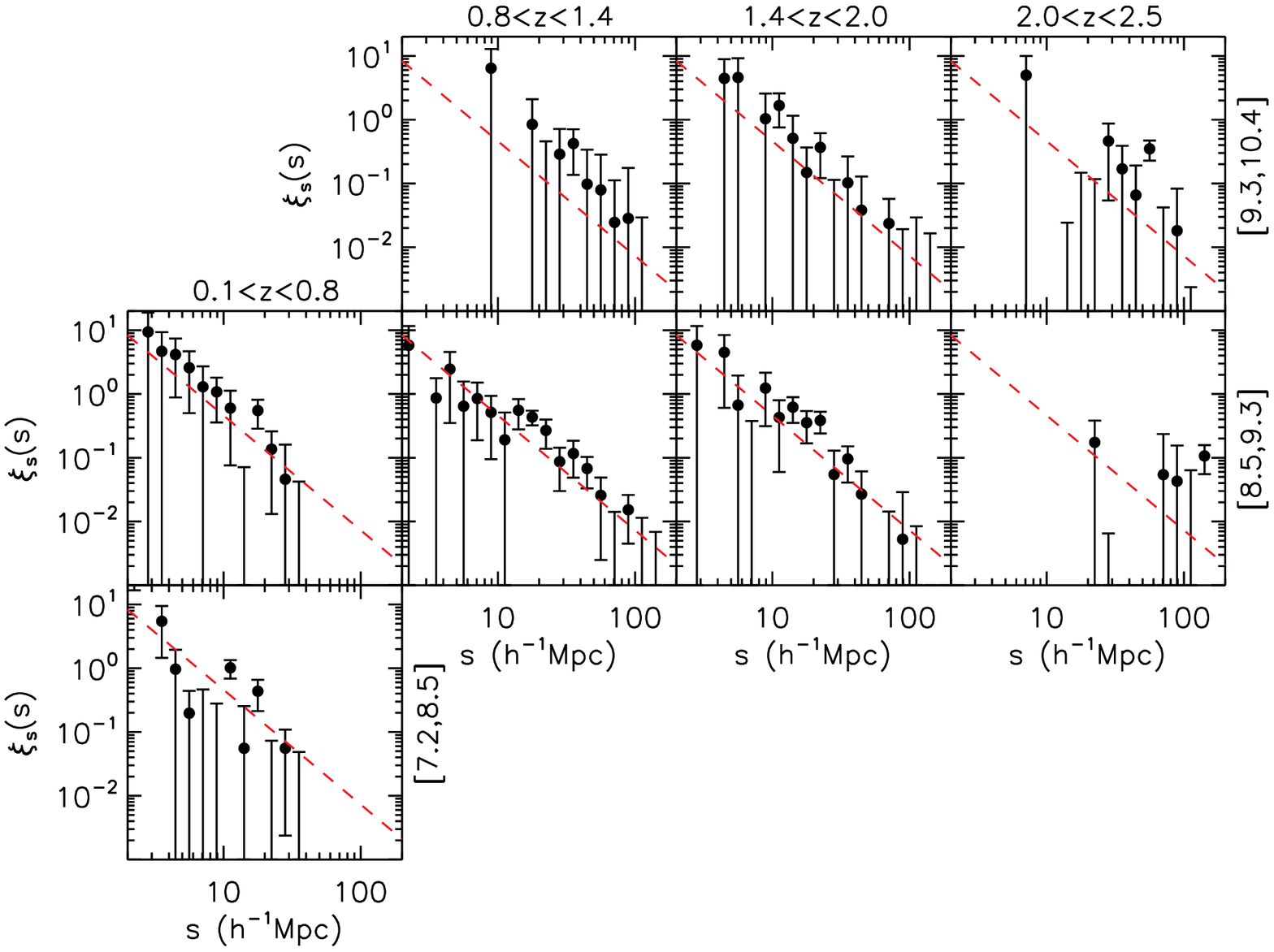}
    \includegraphics[width=0.7\textwidth]{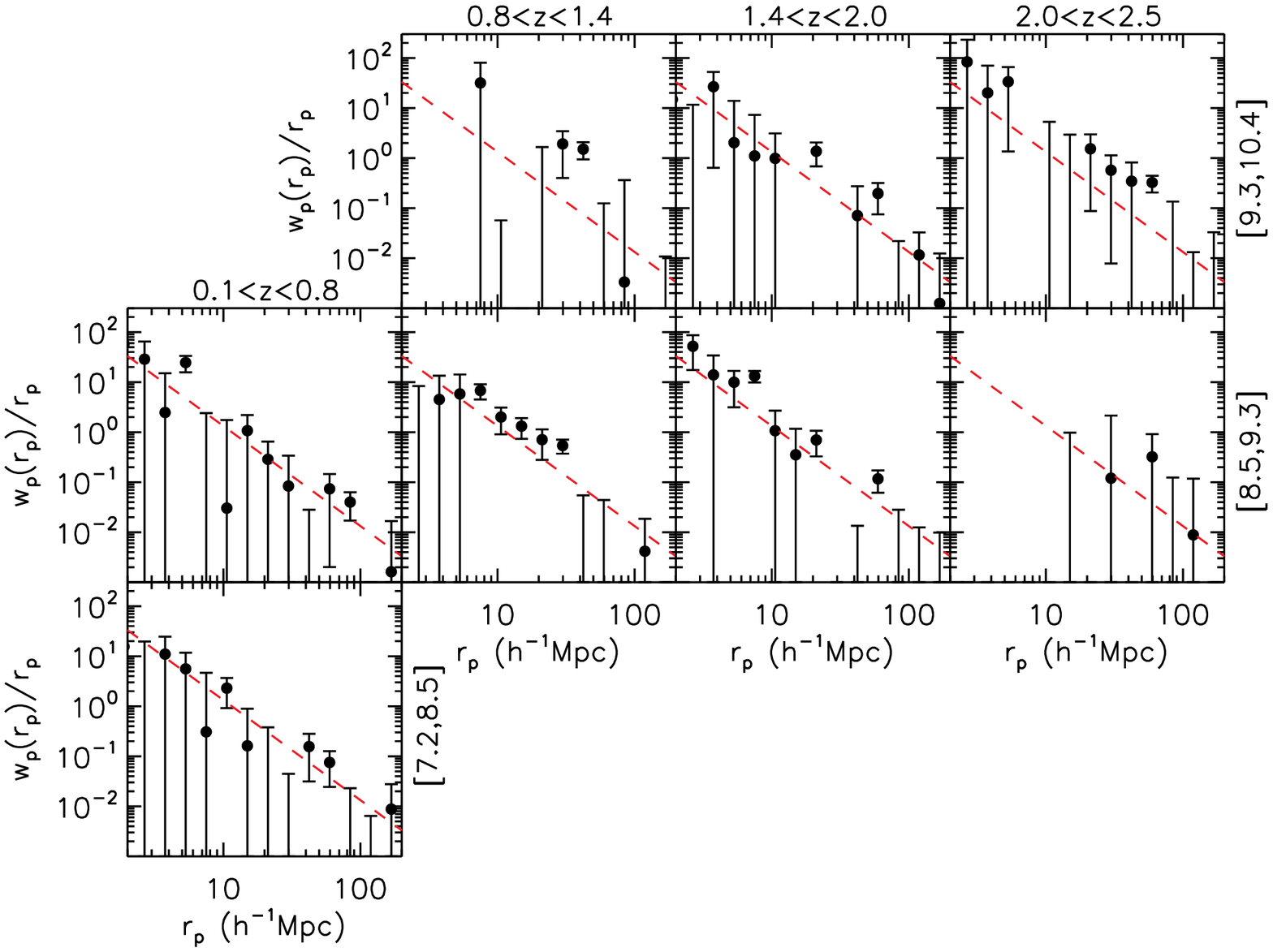}
    \caption{Correlation functions in each redshift-virial mass bin as defined in
    the right panel of Fig. \ref{fig:z_L_BH_div}. luminosity bin as defined
    in the left panel of Fig. \ref{fig:z_L_BH_div}. The redshift
    ranges are marked on the top of each column and
    the virial mass $(\log M_{\rm BH,vir}/M_\odot)$ ranges
    are marked on the right of each row. {\em Upper:} redshift space
    correlation function. {\em Bottom:} projected correlation function. In each
    panel, the dashed line is a power law model of $\xi_s(s)=(s/6.5\ h^{-1}{\rm Mpc})^{-1.8}$ or
    $\xi(r)=(r/6.5\ h^{-1}{\rm Mpc})^{-2}$, drawn
    to guide the eye. No appreciable
    difference is observed at fixed redshift, due to the large uncertainties in the
    correlation functions.}
    \label{fig:BHz_fine}
\end{figure*}

Similarly we show the comparison between virial mass bins at each
redshift bin (as indicated in the right panel of Fig.
\ref{fig:z_L_BH_div}) in Fig. \ref{fig:BHz_fine}. Again no
appreciable mass dependence is observed, although the S/N is poor
in most bins.

\subsubsection{Color Dependence}\label{subsec:color_dep}
Quasars have different colors, either intrinsic or dust reddened.
As discussed in \S\ref{subsec:sample}, we have followed the
approach of Richards et al. (2003) to divide our quasar sample by
median $g-i$ color excess $\Delta(g-i)$ for $0.4\le z\le 2.5$,
which removes the mean color at each redshift [e.g., division
({\em c}) in \S\ref{subsec:sample}; see Fig. \ref{fig:color_div}].

\begin{figure*}
  \centering
    \includegraphics[width=0.45\textwidth]{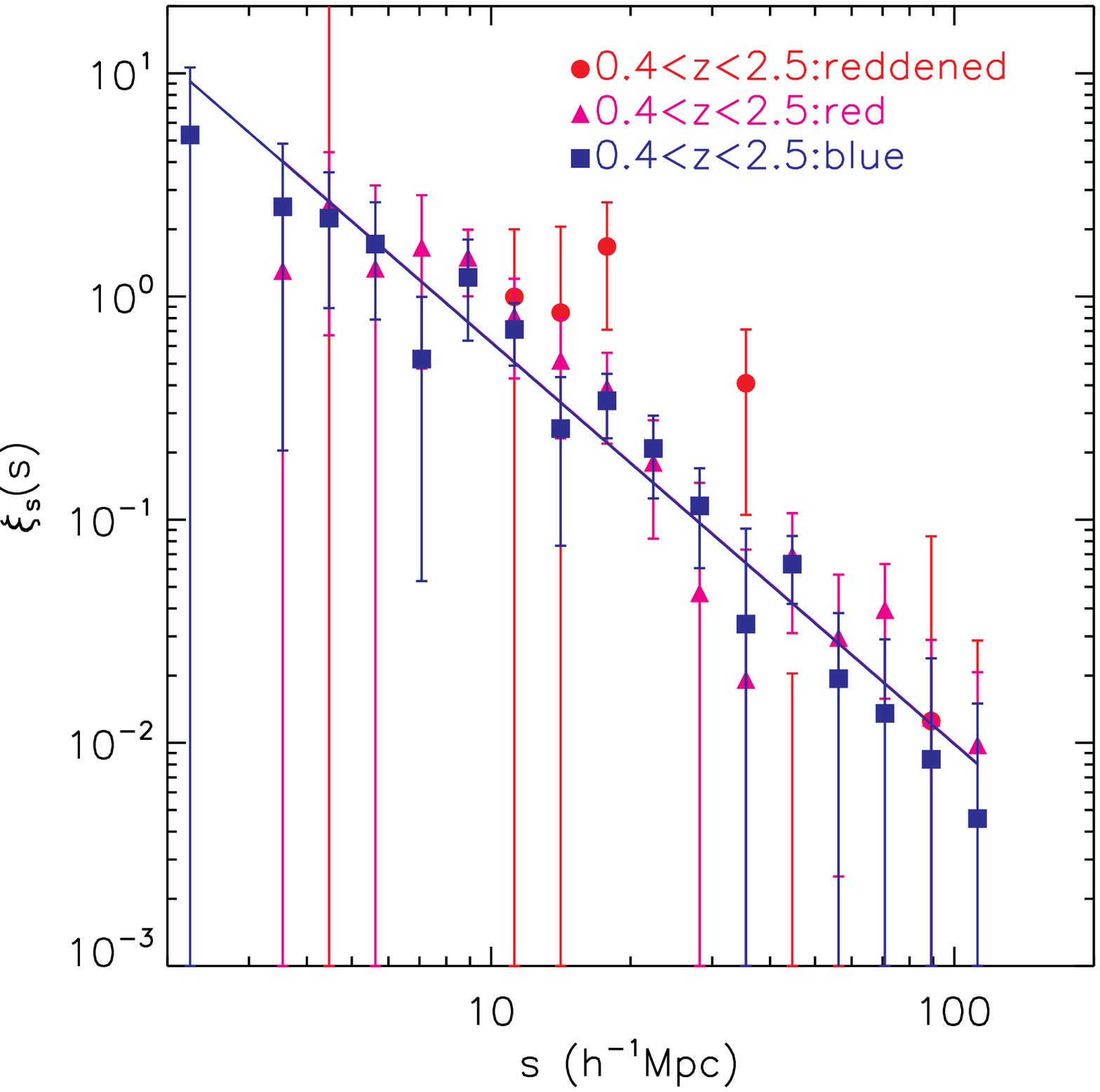}
    \includegraphics[width=0.45\textwidth]{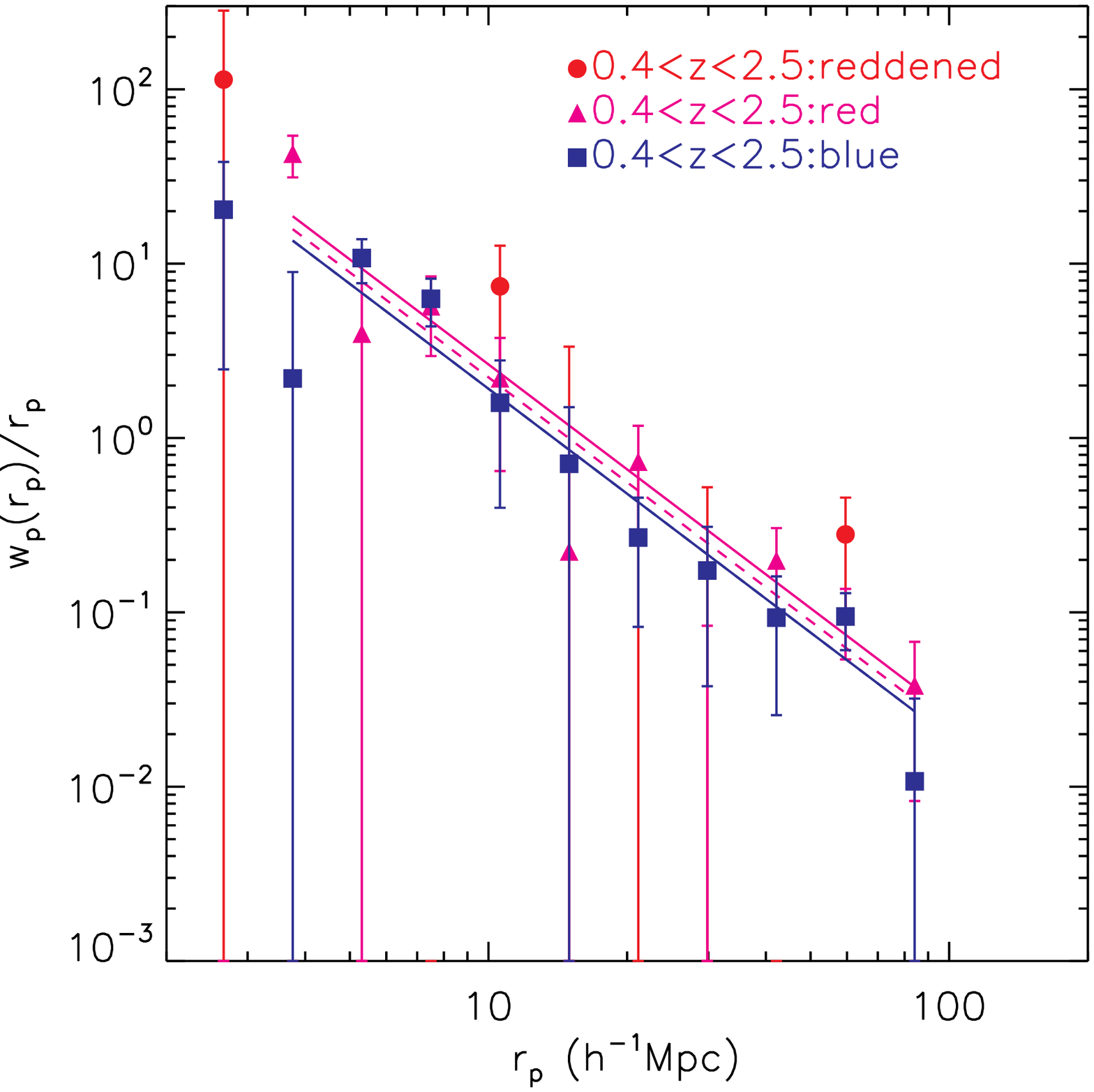}
    \includegraphics[width=0.45\textwidth]{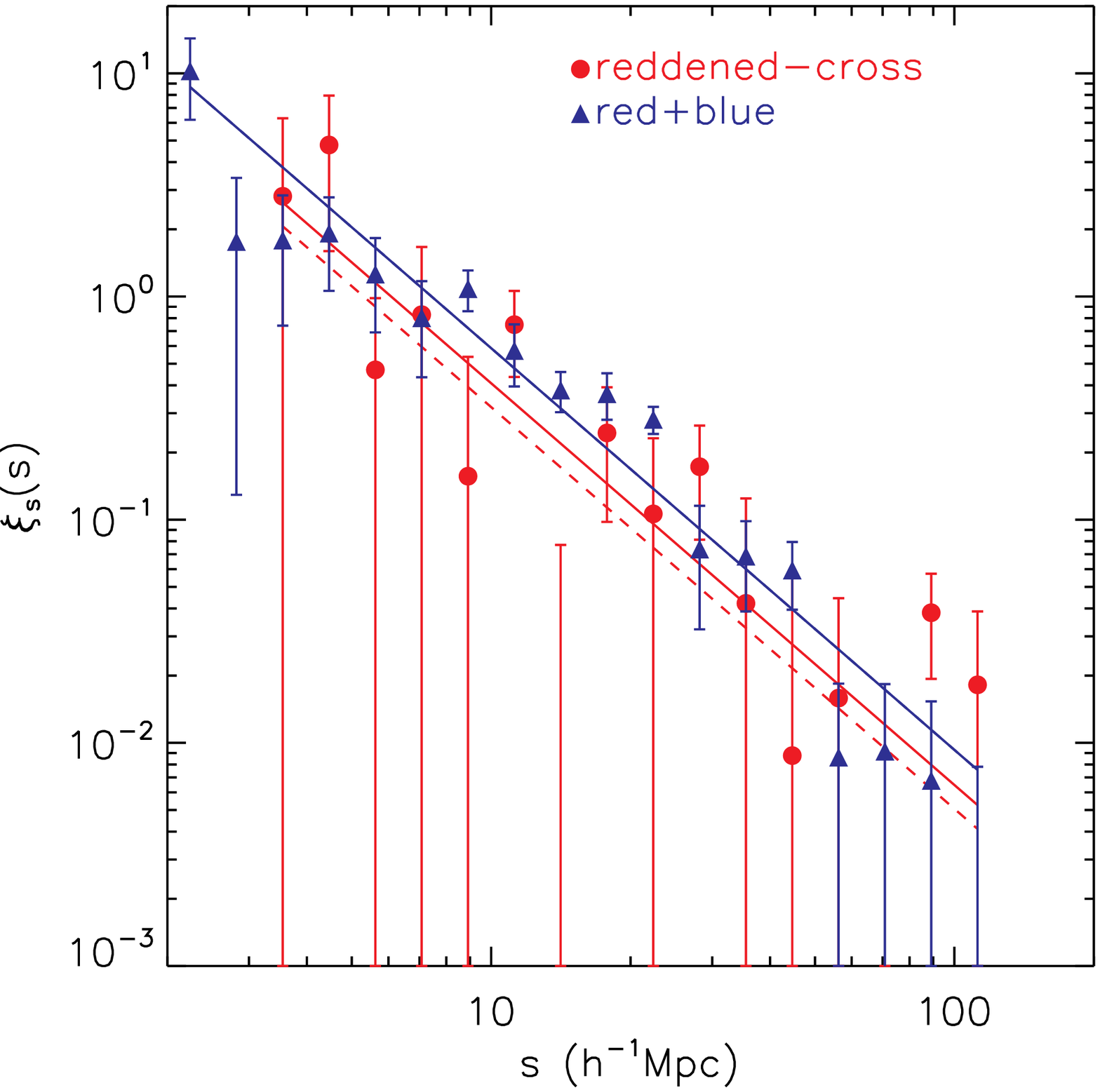}
    \includegraphics[width=0.45\textwidth]{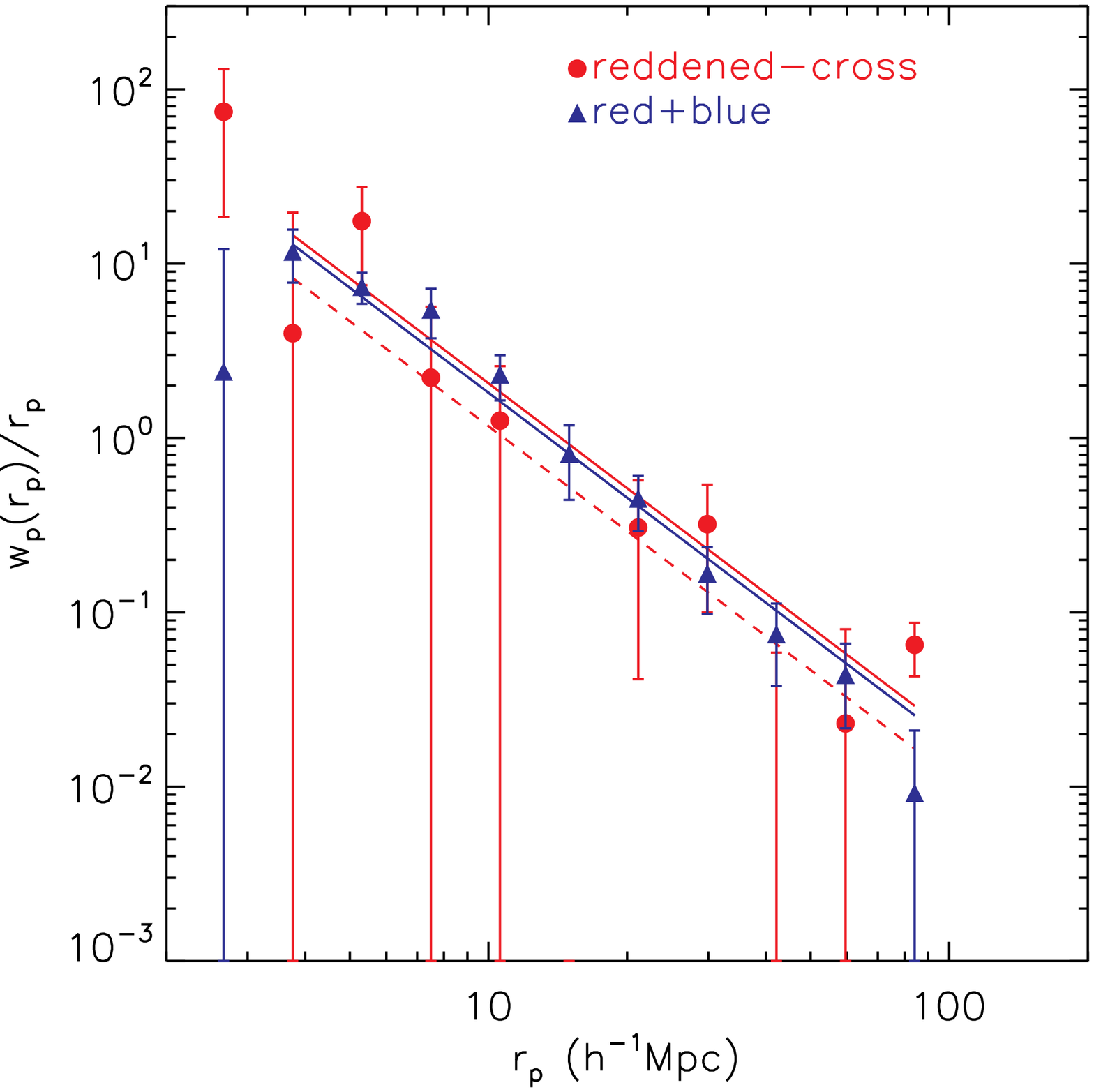}
    \caption{Clustering of blue, red and reddened quasars, using the samples defined in
    \S\ref{subsec:sample}. Solid and dashed lines are the best-fit
    power-law models excluding and including negative data bins
    in the fits, respectively. {\em Left panels:} redshift space
    correlation functions. {\em Right panels:} projected correlation functions. In the upper
    two panels we show the auto-correlation functions for the three samples; there is no
    appreciable difference between the red and blue quasar samples. The auto-correlation function
    of the reddened sample is too noisy to be useful. In the bottom panels, we show the auto-correlation
    functions for the red$+$blue sample and the cross-correlation of the reddened sample with the red$+$blue sample.
    Again no discernible difference is observed given the uncertainty levels in the CF measurements.}
    \label{fig:color_dep}
\end{figure*}

The red and blue quasar samples are sufficiently large to perform
auto-correlations, and we show the results in the upper two panels
of Fig. \ref{fig:color_dep}. The red and blue quasars are
consistent with having similar clustering properties. The reddened
quasar sample is too sparse to do auto-correlation, so we
cross-correlate the blue$+$red quasar sample with the reddened
quasar sample, and show the results in the bottom two panels of
Fig. \ref{fig:color_dep}. The reddened quasars have similar
clustering amplitude. Thus our data indicate there is weak or no
dependence of quasar clustering on colors.

\subsubsection{Radio Dependence}\label{subsec:radio_dep}
\begin{figure*}
  \centering
    \includegraphics[width=0.45\textwidth]{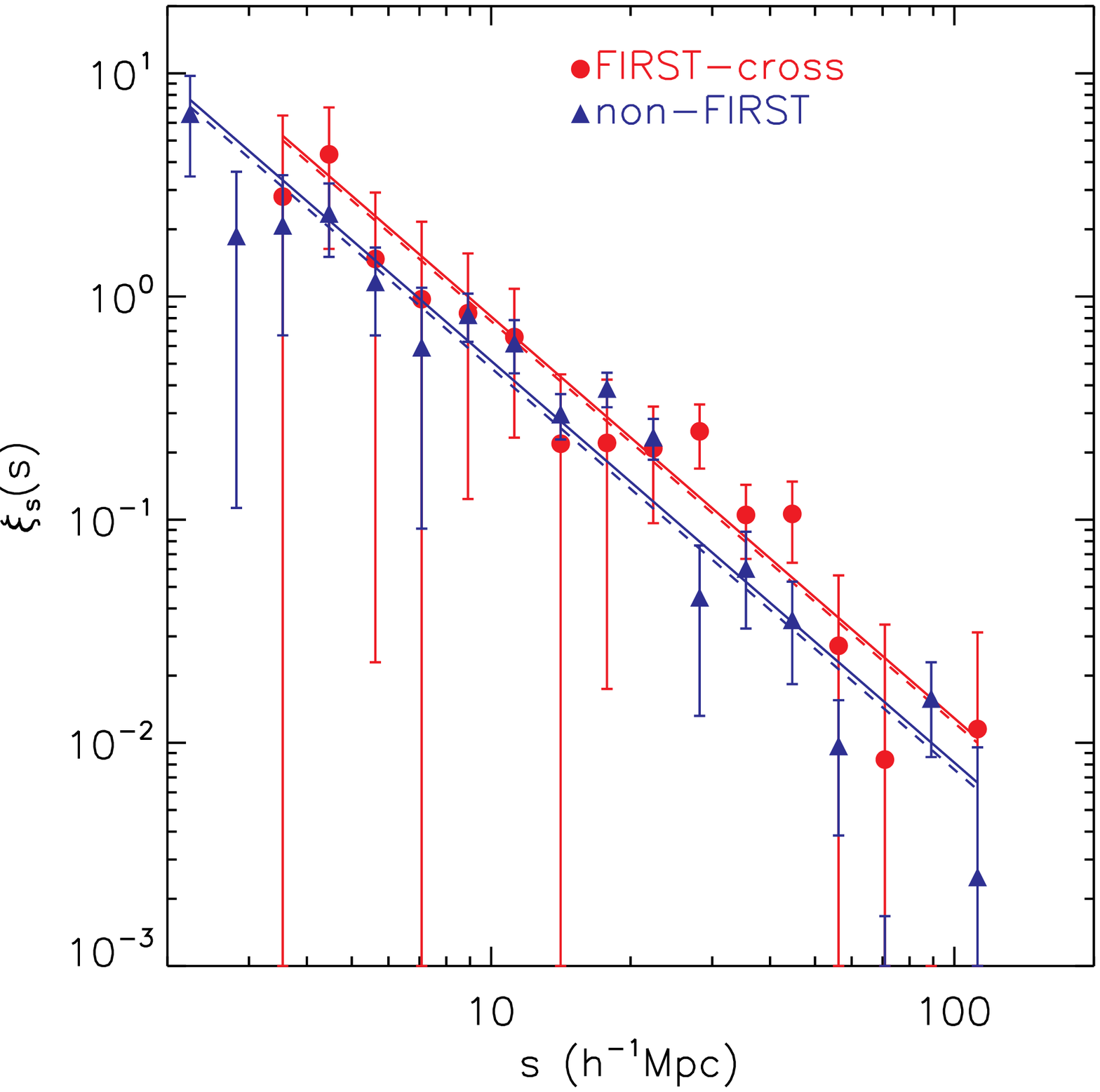}
    \includegraphics[width=0.45\textwidth]{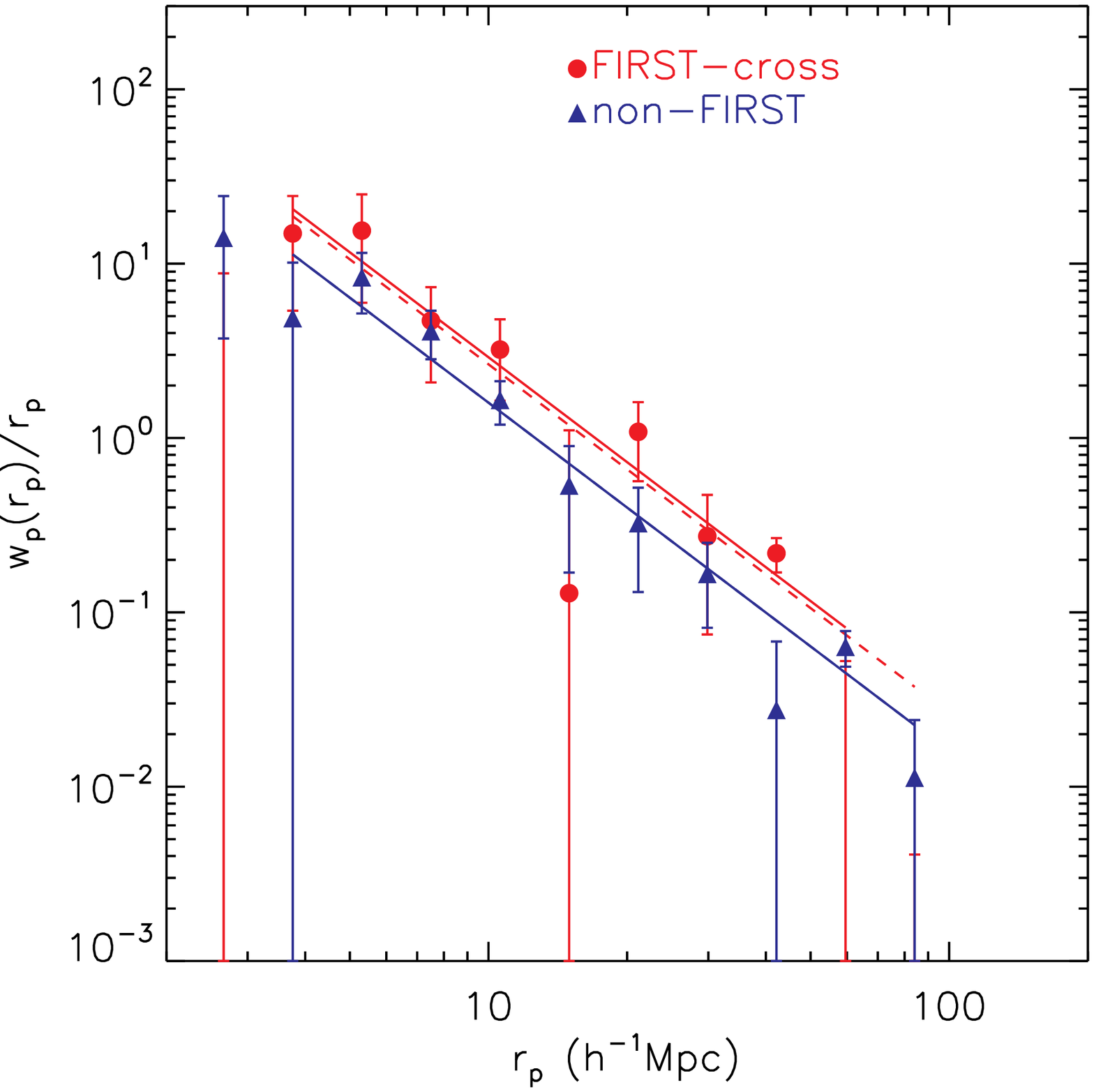}
    \caption{Clustering results for the radio-loud and radio-quiet samples as defined in
    \S\ref{subsec:sample}. Solid and dashed lines are the best-fit
    power-law models excluding and including negative data bins
    in the fits, respectively.
    {\em Left:} redshift space correlation function. {\em Right:} projected correlation function.
     The cross-correlation function of the radio-loud sample with the radio-quiet sample is larger than
     the auto correlation function of the latter sample, as shown in both panels.}
    \label{fig:radio_dep}
\end{figure*}

Finally we investigate if there is any dependence of quasar
clustering on radio activity. Using our $0.4\le z\le 2.5$
FIRST-detected and undetected samples as indicated in Fig.
\ref{fig:radio_div}, we show their clustering comparison in Fig.
\ref{fig:radio_dep}, where we use auto-correlation for the
FIRST-undetected quasars and cross-correlate them with the
FIRST-detected quasars. The FIRST-detected quasars are appreciably
more strongly clustered (cross correlation length $r_0=9.61\pm
0.85\ h^{-1}{\rm Mpc}$) than the FIRST-undetected quasars
($r_0=7.12\pm0.50\ h^{-1}{\rm Mpc}$), which indicates that radio
quasars live in more massive dark matter halos. These
FIRST-detected quasars also tend to have systematically larger
virial BH masses than those FIRST-undetected quasars by $0.12$ dex
(e.g., Shen et al. 2008b). However, the difference in their
clustering remains when we consider a radio-undetected quasar
sample with virial masses matched to those of the radio sample.
Thus more massive host halos, and denser environments may be
related to the triggering of radio activity.

\subsection{A Maximum-Likelihood Approach}\label{subsec:ML_CF}
At the smallest separations, there is only a handful of pairs
distributed among the bins, hence binning effects may become
important, and the usage of the projected correlation function
$w_p(r_p)$ becomes questionable. Here we design a
Maximum-Likelihood (ML) method to use in this situation. Poisson
statistics are assumed to apply, hence we only focus on the
smallest bins where the pairs are statistically independent, i.e.,
the total number of pairs within each bin is much less than the
total number of quasars in the sample.

We take a power-law model for the underlying real space
correlation function $\xi(r)\equiv (r/r_{0,{\rm
ML}})^{-\gamma_{\rm ML}}$. Then we compute the expected numbers of
quasar-quasar and random-quasar pairs for a specified quasar
sample, within a comoving cylindrical volume with projected radius
$R$ to $R+dR$ and half-height $\Delta H=30\ h^{-1}$Mpc; the usage
of a cylindrical volume is to minimize the effects of redshift
distortions and errors, and we found our results are insensitive
to the value of $\Delta H$. In mathematical form, the expected
number of quasar pairs within the cylindrical annulus volume is:
\begin{equation}
\mu= 2\pi R\bar{g}(R)dR\int_{-\Delta H}^{\Delta
H}[1+\xi(r)]dH\times N_{Q}\equiv 2\pi Rh(R)dR\ ,
\end{equation}
where $r\equiv (R^2+H^2)^{1/2}$, $N_{Q}$ is the number of quasars
included in the sample under consideration, and $2\pi
R\bar{g}(R)dR$ is the number of random-quasar pairs within
interval $dR$ divided by the height $2\Delta H$, averaged over
$N_{Q}$ quasars. To compute\footnote{Note that in computing
$\bar{g}(R)$ we have divided by a factor of 2 because we are
calculating the number of pairs, not the number of companions.}
$\bar{g}(R)$ we have used the number density of quasars of the
sample under study as a function of redshift and the polygons
which define the angular geometry of our sample (see appendix B of
Shen et al. 2007a; and Hamilton \& Tegmark 2004 for the
description of spherical polygons).

Following Croft et al. (1997, also see Stephens et al. 1997) we
write the likelihood function as
\begin{equation}
{\cal L}=\prod_i^N e^{-\mu_i}\mu_i\prod_{j\neq i}e^{-\mu_j}\ ,
\end{equation}
where $\mu_i=2\pi R_ih(R_i)dR$, the expected number of pairs in
the interval $dR$, and the index $j$ runs over all the elements
$dR$ in which there are no pairs. Defining the usual quantity
$S\equiv -2\ln{\cal L}$ we have
\begin{equation}
S \equiv -2\ln {\cal L} = 2\int_{R_{\rm min}}^{R_{\rm max}} 2\pi
Rh(R)dR - 2\sum_i^N\ln[h(R_i)]\ ,
\end{equation}
with all the model independent terms removed. Here the summation
is over all $N$ quasar pairs we found in our sample, and $[R_{\rm
min},R_{\rm max}]$ is the range of scales over which we search for
quasar pairs. The values of $R_{\rm min}$ and $R_{\rm max}$ are
chosen to avoid fiber collision effects and to ensure that the
pairs found are statistically independent, i.e., $R_{\rm
min}\gtrsim 2\ h^{-1}$Mpc and generally $R_{\rm max}\lesssim 20\
h^{-1}$Mpc. But we find that different ranges of $[R_{\rm
min},R_{\rm max}]$ do yield different results, which do not always
overlap within 1$\sigma$. Hence we examine different ranges of
$[R_{\rm min},R_{\rm max}]$ for each sample. We then minimize $S$
with respect to $r_{0,{\rm ML}}$ and obtain its $1\sigma$
uncertainty. Again, because of the noisy data we fix $\gamma_{\rm
ML}=2$ in our fitting procedure.

Compared with binned CFs, the ML procedure has several drawbacks.
The information from pairs with large separations ($R>R_{\rm
max}$) is not used, and the merits of the ML method suffer if
there is a paucity of pairs due to systematic effects (see below).
For these reasons, we only apply this ML method and compare the ML
estimates with our binned CF results as a consistency check (see
Table \ref{table:CF_result}).

\begin{figure}
  \centering
    \includegraphics[width=0.45\textwidth]{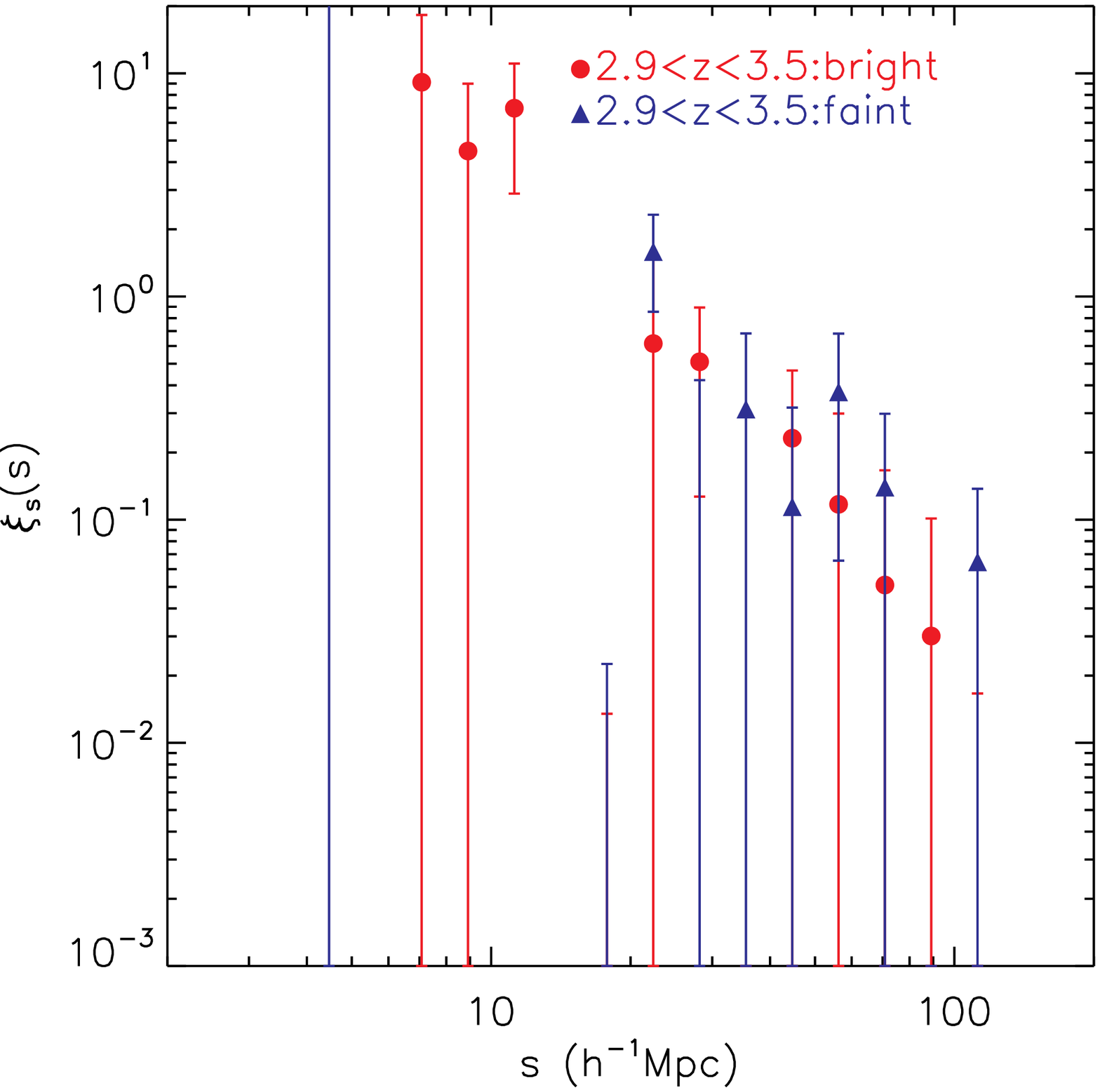}
    \includegraphics[width=0.45\textwidth]{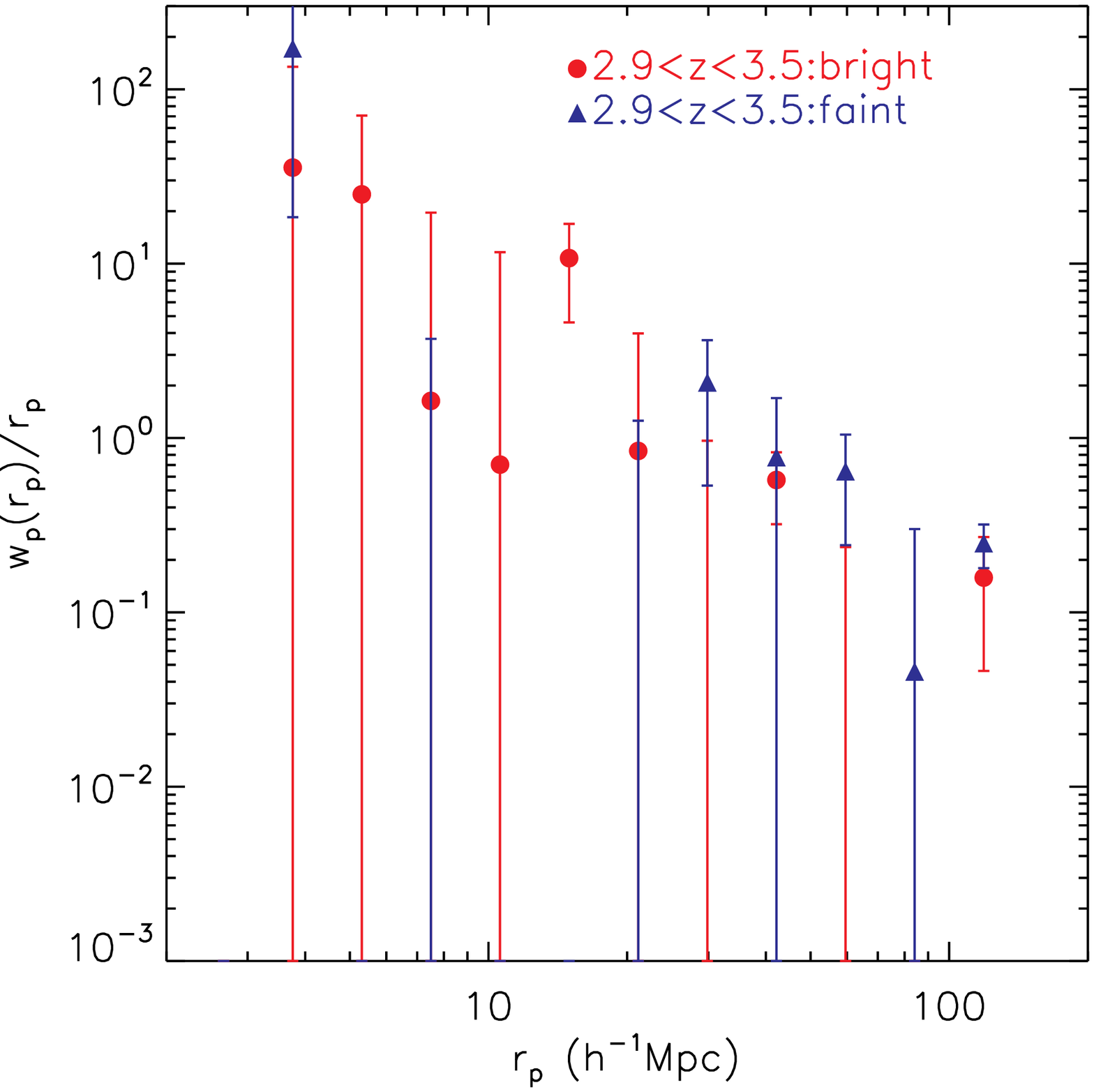}
    \caption{Binned CFs for the two high-$z$ luminosity bins. {\em Left:} redshift space
    correlation function. {\em Right:} projected correlation function. Despite the
    absence of detectable correlation for the fainter sample (triangles) at $r\lesssim 20\ h^{-1}$Mpc, which
    may be caused by systematic effects (see the text), there is no appreciable
    difference at large scales ($r\gtrsim 20\ h^{-1}$Mpc) for the two samples, although
    the error bars are large.}
    \label{fig:highz_Ldep}
\end{figure}

We start with the two luminosity bins at $2.9<z<3.5$ shown in the
left panel of Fig. \ref{fig:z_L_BH_div}. In Fig.
\ref{fig:highz_Ldep} we show the binned CFs for these two
luminosity bins. On small scales ($r_p\lesssim 15\ h^{-1}$Mpc), an
apparent deficit of correlation is seen for the fainter bin with
respect to the brighter bin. On larger scales, the signal-to-noise
ratios of the CFs are too small to distinguish any difference in
clustering strength.
For the ML approach we choose a comoving volume of a cylinder
annulus with half-height $\Delta H=30\ h^{-1}$Mpc and $[R_{\rm
min},R_{\rm max}]=[2,15]\ h^{-1}$Mpc, $[5,20]\ h^{-1}$Mpc, and
$[20,40]\ h^{-1}$Mpc. The best fit ML correlation lengths are
listed in Table \ref{table:CF_result}, which show no detectable
correlations at small scales ($r_p\lesssim 15\ h^{-1}$Mpc) for the
fainter bin, consistent with the binned CF results.

The relative paucity of small-separation pairs in the fainter
$2.9<z<3.5$ luminosity bin is much stronger than seen in the
$z<2.9$ luminosity bins. However, we suspect this is at least
partly caused by some not well-understood systematic effects in
our sample. The lower redshift cut $z=2.9$ is close to the
redshift desert at $z\sim 2.7$ where the quasar selection
efficiency drops. The completeness becomes worse close to the
magnitude cut $i=20.2$ at $z\sim 2.9$, as seen in our sample (also
see fig. 17 of Richards et al. 2006). As we argued above, image
quality becomes important in selecting quasars close to the faint
magnitude cut, and preferential loss of quasar pairs with similar
colors is possible. Therefore we do not claim that the apparent
weaker clustering of fainter quasars at $2.9<z<3.5$ and at scales
$r\lesssim 20\ h^{-1}$Mpc is significant. On the other hand, our
sample is unable to give clear evidence of a dependence of
clustering on luminosity at large scales at $z>2.9$ because of the
sample sparseness. We hope that ongoing and future surveys of
high-redshift quasars at the faint luminosity end can help resolve
these issues.


We now apply the ML method to several subsamples in different
redshift bins,
and compare the results with those using binned method in Table
\ref{table:CF_result}. The ML method yields results consistent
with the binned method within error bars. In fact, the difference
in the best-fit values of $r_{\rm 0,ML}$ for various fitting
ranges in the ML approach is fully reflected in the binned CF data
points, i.e., we get small values of $r_{\rm 0,ML}$ for fitting
ranges where the binned CF data points fall below the power-law
fit.

\section{DISCUSSION}\label{sec:diss}
\begin{figure*}
  \centering
    \includegraphics[width=0.8\textwidth]{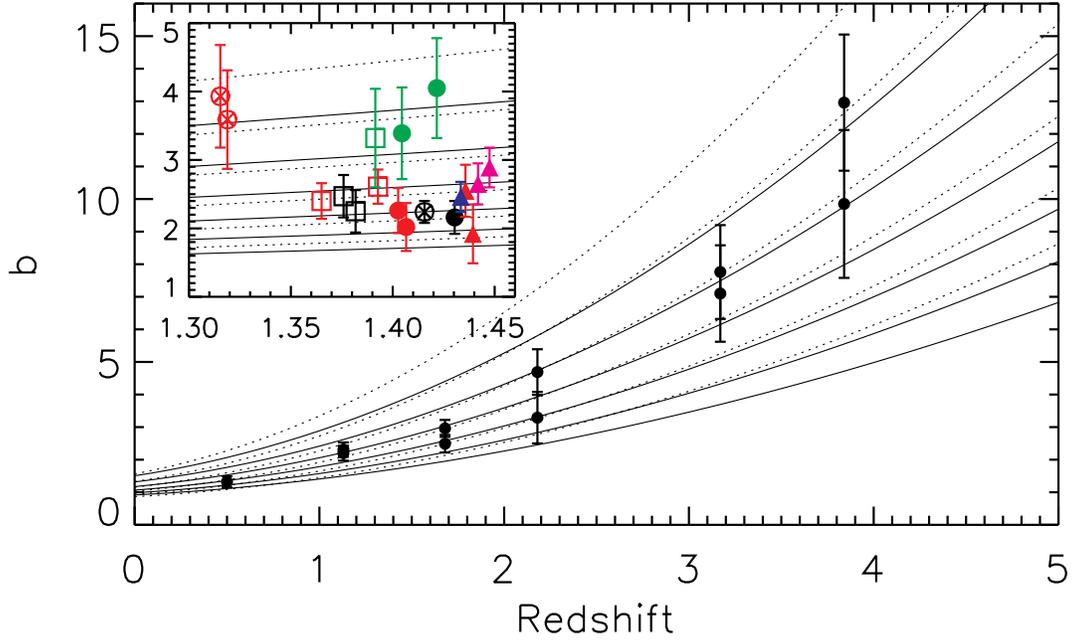}
    \caption{Measured quasar bias factors for various samples. The filled circles in the main
    frame are results for all quasars within each redshift bin (no magnitude cut) and results
    both with and without negative data points in the fitting procedure are shown. The solid and dotted lines
    are halo bias factor evolution for fixed halo mass (from bottom to top) $M_{\rm halo}=5\times10^{11},\
    1\times10^{12},\ 2\times10^{12},\ 4\times10^{12},\ 8\times10^{12},\ 1.6\times10^{13}\ h^{-1}M_\odot$
    using the Sheth et al. (2001) and Jing (1998) fitting formulae respectively. The Jing (1998) formula
    generally gives larger halo bias than the Sheth et al. (2001) formula for the same halo mass. In the inset we show
    the quasar biases for our $0.4<z<2.5$ assembly samples divided by luminosity ({\em filled circles; black for faint quasars, red for bright quasars, and
    green for the brightest quasars}),
    virial mass ({\em open squares; black for low-mass, red for high-mass and green for the most massive quasars}), color ({\em filled triangles, blue for blue quasars,
    magenta for red quasars and red for reddened quasars}), and radio
    detection
    ({\em circles with x; black for FIRST-undetected quasars and red for FIRST-detected quasars}). Their median redshifts have been shuffled to avoid clustering in the plot.}
    \label{fig:bias_compile}
\end{figure*}

We now discuss the dependences (or lack thereof) of quasar
clustering on various quantities that we have found. Our
clustering measurements of various subsamples are summarized in
Fig. \ref{fig:bias_compile} in terms of the bias factor. The goal
is to compare the observed clustering measurements with
theoretical predictions; or alternatively, to use these
measurements to constrain theoretical models, i.e., the form and
scatter of various relations, such as relations of true BH mass
versus host halo mass, luminosity versus true BH mass/host halo
mass, and virial BH mass versus true BH mass, etc. We assume that
halo mass is the only relevant quantity that correlates with
clustering strength. Hence the question in any theoretical model
is how to populate halos with quasars of various physical
properties (luminosity, BH mass etc), given constraints from the
halo abundance, the quasar luminosity function and quasar
clustering.


\subsection{Luminosity versus Host Halo Mass}\label{subsec:L_vs_Mhalo}
In this subsection we outline the basic logic in modeling the
luminosity dependence of quasar clustering. Suppose that at fixed
host halo mass, the quasar luminosity follows a correlation given
by:
\begin{equation}\label{eqn:L_M}
\log L=C + \alpha\log M_{\rm halo} + \sigma_L\ ,
\end{equation}
where $C$ and $\alpha$ are constants, and $\sigma_L$ is a Gaussian
random deviate. For simplicity we have assumed here a power-law
form with log-normal scatter, but more complex correlations are
possible. This correlation (\ref{eqn:L_M}) gives the probability
distribution of ${\cal P}(L|M_{\rm halo})$ as,
\begin{equation}
{\cal P}(L|M_{\rm
halo})=(2\pi\sigma_L^2)^{-1/2}\exp\bigg\{-\frac{[\log L-(C +
\alpha\log M_{\rm halo})]^2}{2\sigma_L^2}\bigg\}\ .
\end{equation}
Using Bayes's theorem we can derive the probability distribution
of host halo mass at fixed luminosity ${\cal P}(M_{\rm halo}|L)$,
given our knowledge of the halo mass function $n(M_{\rm halo})$
and the halo duty cycle $f_{\rm on}$, the probability of a halo
hosting an active quasar:
\begin{equation}\label{eqn:prob_M}
{\cal P}(M_{\rm halo}|L)\propto {\cal P}(L|M_{\rm halo})n(M_{\rm
halo})f_{\rm on}(M_{\rm halo})\ ,
\end{equation}
where the halo mass function $n(M_{\rm halo})$ is usually
extracted from simulations. Generally there will be a
Malmquist-type bias because of the scatter $\sigma_L$ and the
bottom-heavy halo mass function. But the exact magnitude of this
bias as function of luminosity also depends on the halo duty cycle
$f_{\rm on}$, which itself may well be a function of halo mass and
redshift, and is constrained via the quasar luminosity function:
\begin{equation}\label{eqn:LF}
\frac{d\Phi}{dL}=\int dM_{\rm halo}{\cal P}(L|M_{\rm
halo})n(M_{\rm halo})f_{\rm on}(M_{\rm halo})\ .
\end{equation}
Therefore given the assumed correlation Eqn. (\ref{eqn:L_M}), halo
mass function, and $f_{\rm on}$ constrained from the quasar
luminosity function, we obtain the mass distribution of halos
hosting quasars shining at any fixed luminosity $L$ via eqn.
(\ref{eqn:prob_M}).

We also must know the linear halo bias factor $b_{\rm halo}$ as
function of halo mass in order to make clustering predictions.
This is somewhat problematic, as noted in White et al. (2008).
Various fitting functions for the halo bias exist in the
literature, which differ in both magnitude and shape (e.g., Mo \&
White 1996; Jing 1998; Sheth \& Tormen 1999; Sheth et al. 2001;
Tinker et al. 2005; Basilakos et al. 2008). In practice one needs
a halo bias formula that is tested against large numerical
simulation sets at the ranges of halo mass and redshift of
interest (e.g., see the discussion in Shankar et al. 2008b).
Finally the effective bias factor of quasars at fixed luminosity
is the average over all the halos that host them:
\begin{equation}
b_L(L)=\int dM_{\rm halo} b_{\rm halo}(M_{\rm halo}){\cal
P}(M_{\rm halo}|L)\ ,
\end{equation}
which is to be compared with observations.

The correlation in eqn. (\ref{eqn:L_M}) is analogous to a light
bulb model in which a quasar shines at a more or less fixed
luminosity (i.e., a fraction of the Eddington luminosity of its
SMBH) when it is on, with a log-normal scatter $\sigma_L$ in
luminosity. Theoretical models predict a scaling $L\propto M_{\rm
halo}^{5/3}$ or $L\propto M_{\rm halo}^{4/3}$ depending on energy
conservation (e.g., Silk \& Rees 1998; Wyithe \& Loeb 2002) or
momentum conservation (e.g., King 2003) during self-regulated
black hole growth. Recent measurements of the $M_{\rm BH}-\sigma$
relation (e.g., Tremaine et al. 2002) favor the latter scaling,
$L\propto M_{\rm halo}^{4/3}$, as revealed in some hydrodynamic
simulations (e.g., Di Matteo et al. 2005). A more sophisticated,
and perhaps more physically motivated, estimate for ${\cal
P}(L|M_{\rm halo})$ has been put forward by P. Hopkins and
collaborators, based upon simulated quasar light curves and
correlations between quasar peak luminosity and halo mass (e.g.,
Hopkins et al. 2005; Lidz et al. 2006). Simple phenomenological
models, however, such as the one that uses eqn. (\ref{eqn:L_M}),
may fit the observations equally well (e.g., Shankar et al. 2008a;
Croton 2008; Shankar et al. 2008b). A full exploration of these
models is beyond the scope of this paper.

To gain some crude sense of how our measurements compare with
theoretical models, we use the diagnostic fig. 3 in Lidz et al.
(2006), which shows predictions for the quasar bias as a function
of luminosity at $z=2$. Their models use the halo bias formula
from Sheth et al. (2001). The dynamical range in the median
luminosity of each bin probed by our sample is quite narrow, $\sim
0.4$ dex, compared to $\sim 1$ dex for the 2QZ+2SLAQ sample used
in da \^{A}ngela et al. (2008). For our bright and faint quasar
samples as defined in \S\ref{subsec:sample} and in the upper-left
panel of Fig. \ref{fig:median_L_BH_div}, the median redshift is
$\bar{z}=1.4$, and the median luminosities are $\bar{L}_{\rm bol}=
10^{46.56}\ {\rm erg\,s^{-1}}$ and $10^{46.31}\ {\rm
erg\,s^{-1}}$. Note that $L_{\rm bol}=10^{46.4}\ {\rm
erg\,s^{-1}}$ corresponds to {\em rest-frame} $B-$band luminosity
$L_{B}\sim 10^{12} L_{\odot}$ in Lidz et al.\ (2006). Below and
around this luminosity, even the light bulb model in their fig. 3
(without any scatter $\sigma$) gives weak luminosity dependence of
quasar clustering. This is consistent with our measurements, as
well as da \^{A}ngela et al. (2008), who probe only a $\lesssim 1$
dex dynamical range in luminosity at redshift below $\sim 2.5$. It
is also marginally consistent with the results at $z\sim 2.5$ in
Adelberger \& Steidel (2005), whose dynamical range in AGN
luminosity is $\sim 2.5$ dex.

When the scatter $\sigma_L$ in eqn. (\ref{eqn:L_M}) is
incorporated, the weighted bias factor at fixed luminosity will
drop due to the scatter from more abundant less massive, and less
biased halos, but it can be fine-tuned to yield an even flatter
luminosity dependence if, for instance, the halo duty cycle
$f_{\rm on}(M_{\rm halo})$ decreases at an appropriate rate when
$M_{\rm halo}$ decreases. One must also correct the overall linear
bias of quasars by adjusting the normalization of the correlation
between luminosity and halo mass (Eqn. \ref{eqn:L_M}), using
constraints from the observed quasar luminosity function and the
theoretical halo mass function, following the logical flow we just
formulated above. At this stage, the majority of the measurements
of the luminosity dependence of quasar clustering for moderate
luminosities are consistent with theoretical predictions of weak
dependence, but are unable to distinguish different models.
Uncertainties in the quasar luminosity function, halo bias factor,
and halo mass function, which are all needed to make a
self-consistent model, further obscure the situation.

Since the halo bias factor increases very rapidly with mass,
various models are sensitive at the high-luminosity end of quasar
clustering. Unfortunately, previous studies were based on samples
containing too few bright quasars to perform correlation analyses;
cross-correlating those bright quasars with much larger galaxy
samples is a promising approach (e.g., Adelberger \& Steidel 2005;
Padmanabhan et al.\ 2008), but statistically significant results
are yet to come. We have shown for the first time that the most
luminous quasars indeed are more strongly clustered than their
fainter counterparts. The median luminosity for our top $10\%$
most luminous quasar sample is $\bar{L}_{\rm bol}=10^{46.84}\ {\rm
erg\,s^{-1}}$, a factor of $\sim 0.5$ dex larger than the median
luminosity for the remaining quasars. With this increment in
luminosity, we are probing the rarest and most biased halos. Using
the Sheth et al. (2001) halo bias formula, we estimate a halo mass
$\sim 1-2\times10^{13}\ h^{-1}M_\odot$ for the $10\%$ most
luminous quasars and $\sim 2\times 10^{12}\ h^{-1}M_\odot$ for the
remainder.

\subsection{Virial BH Masses versus Host Halo Mass}\label{subsec:BH_vir_vs_Mhalo}
We may also expect a correlation between host halo mass and black
hole mass. There have been claims for such a correlation at low
redshift, where the circular velocity is used as a surrogate for
halo mass and the central velocity dispersion is used as a
surrogate for BH mass (e.g., Ferrarese 2002; Baes et al. 2003),
although significant uncertainty remains due to limited sample
size and systematic effects; for instance, it is argued that this
correlation is dependent on Hubble type (e.g., Courteau et al.
2007; Ho 2007). But for now let us neglect the systematic effects,
and, analogous to eqn. (\ref{eqn:L_M}), write the correlation
between central BH mass and halo mass as:
\begin{equation}\label{eqn:BH_Mhalo}
\log M_{\rm BH} = C +\beta\log M_{\rm halo}+\sigma_{\rm BH}\ ,
\end{equation}
where the constants $C$ and $\beta$ define the normalization and
scaling of this correlation, and the random Gaussian deviate
$\sigma_{\rm BH}$ denotes the scatter of this correlation.
Neglecting the scatter $\sigma_{\rm BH}$ and taking $\beta\sim
1.65$ as measured in Ferrarese (2002), one dex in $M_{\rm BH}$
corresponds to $\sim 0.6$ dex in $M_{\rm halo}$.

The virial masses we measure for our quasar sample are scattered
from the true masses by $\sim 0.3-0.4$ dex (see discussion in Shen
et al.\ 2008b). To use eqn. (\ref{eqn:BH_Mhalo}) and its inverse
relation we need a model to disentangle the intrinsic BH masses
from the estimated virial BH masses.


Shen et al. (2008b) proposed a simple model of the intrinsic BH
mass and Eddington ratio distributions which simultaneously
reproduces the quasar luminosity and virial BH mass distributions
in different luminosity bins for SDSS quasars. In this model, the
intrinsic BH mass has a power law distribution, and the Eddington
ratio has a log-normal distribution at fixed intrinsic BH mass
with a dispersion $0.4$ dex (see \S\S 4.3 and 4.4 in Shen et al.
2008b for details). We follow the methodology outlined in \S4.3 in
that paper and use the model parameters for the \MgII\ virial mass
relation in their Table 2 to simulate both true and virial BH
masses and bolometric luminosities. These parameters are found to
be redshift-independent (Shen et al. 2008b). We then use equation
(\ref{eqn:Mi_Lbol}) to simulate the $i-$band absolute magnitudes
and impose the SDSS magnitude cut ($i=19.1$ at $0.4\le z\le 2.5$)
until we achieve the same redshift distribution of observed virial
BH masses as shown in Fig. \ref{fig:median_L_BH_div}.

\begin{figure*}
  \centering
    \includegraphics[width=0.45\textwidth]{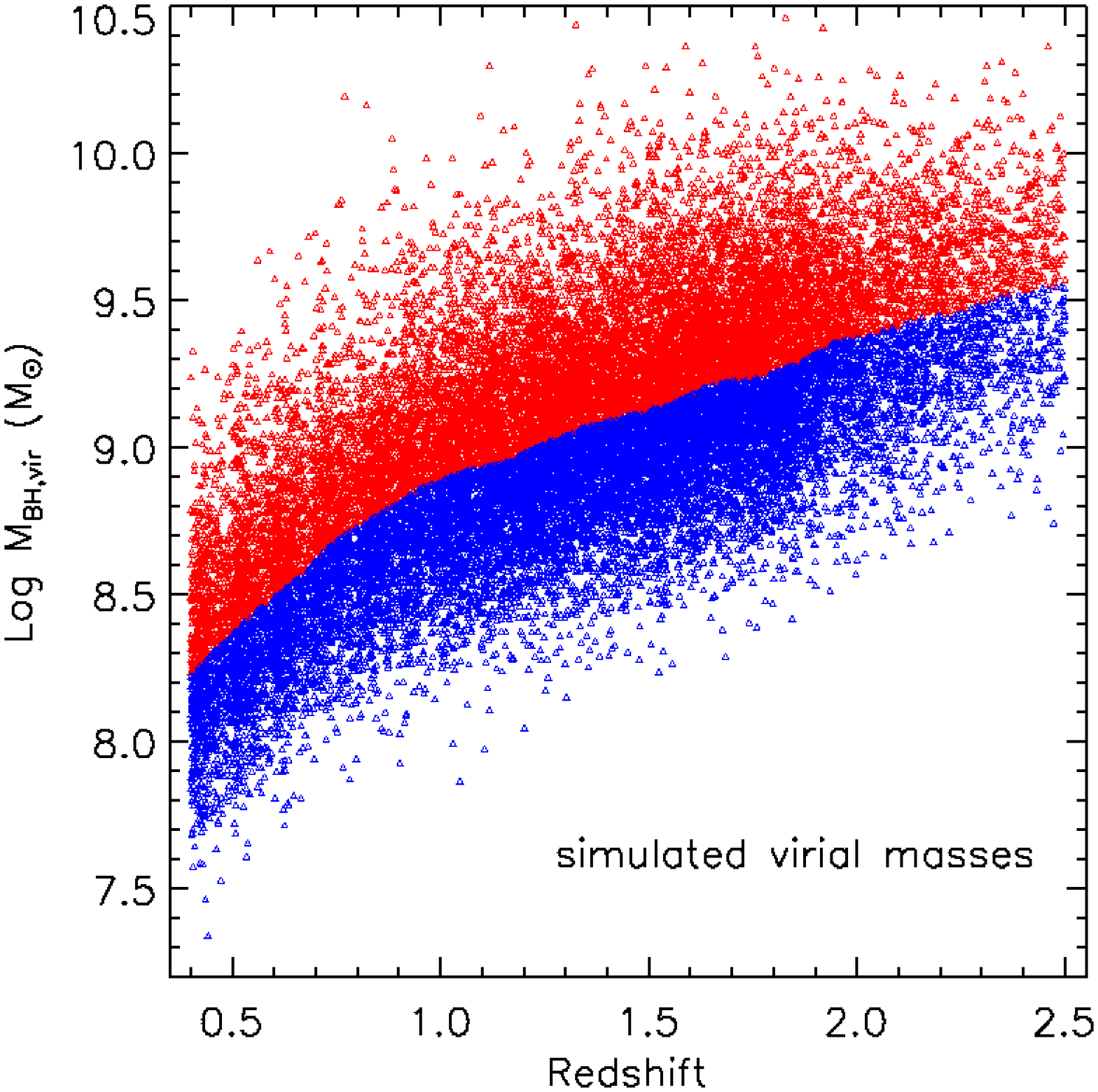}
    \includegraphics[width=0.45\textwidth]{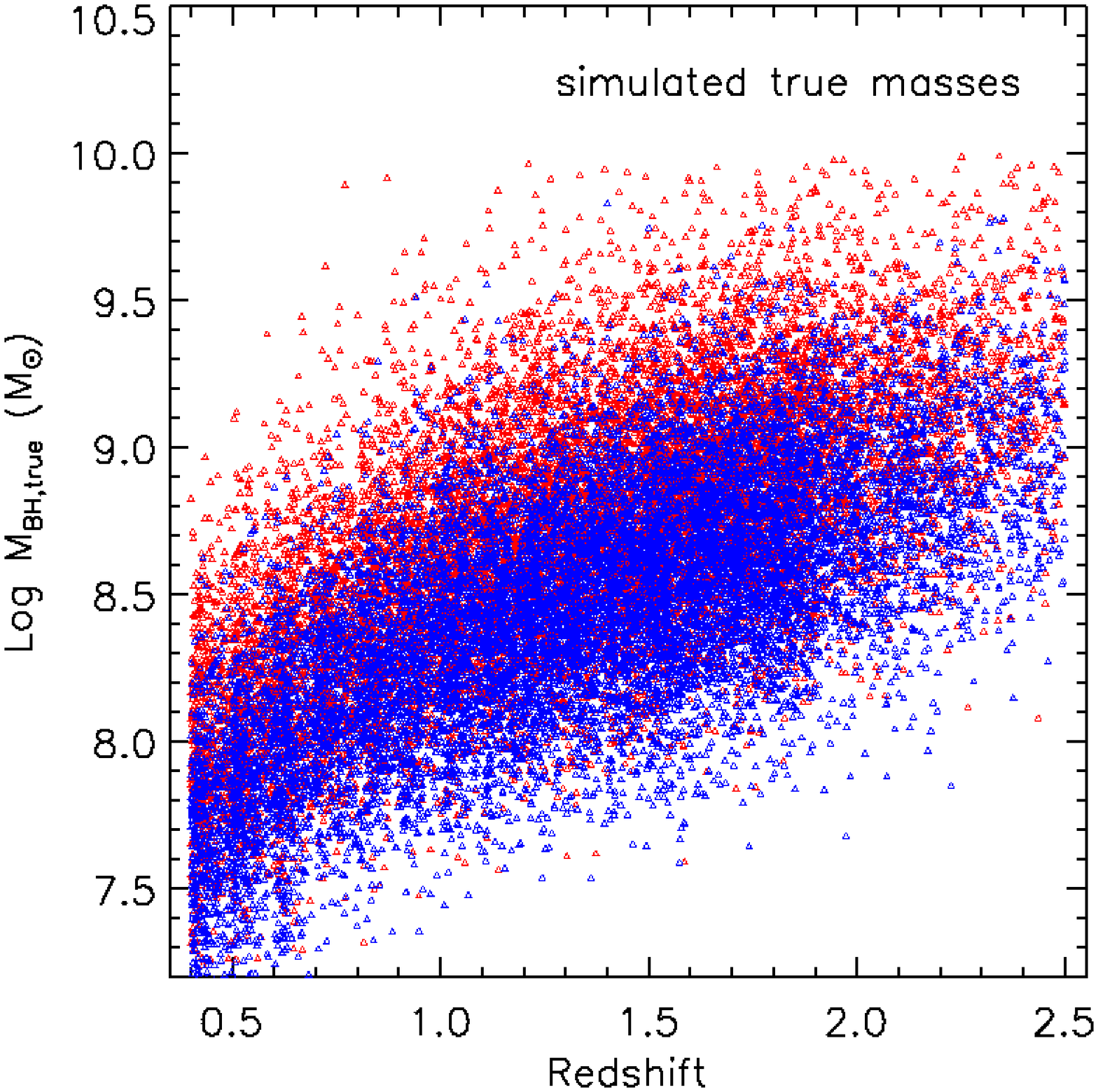}
    \includegraphics[width=0.45\textwidth]{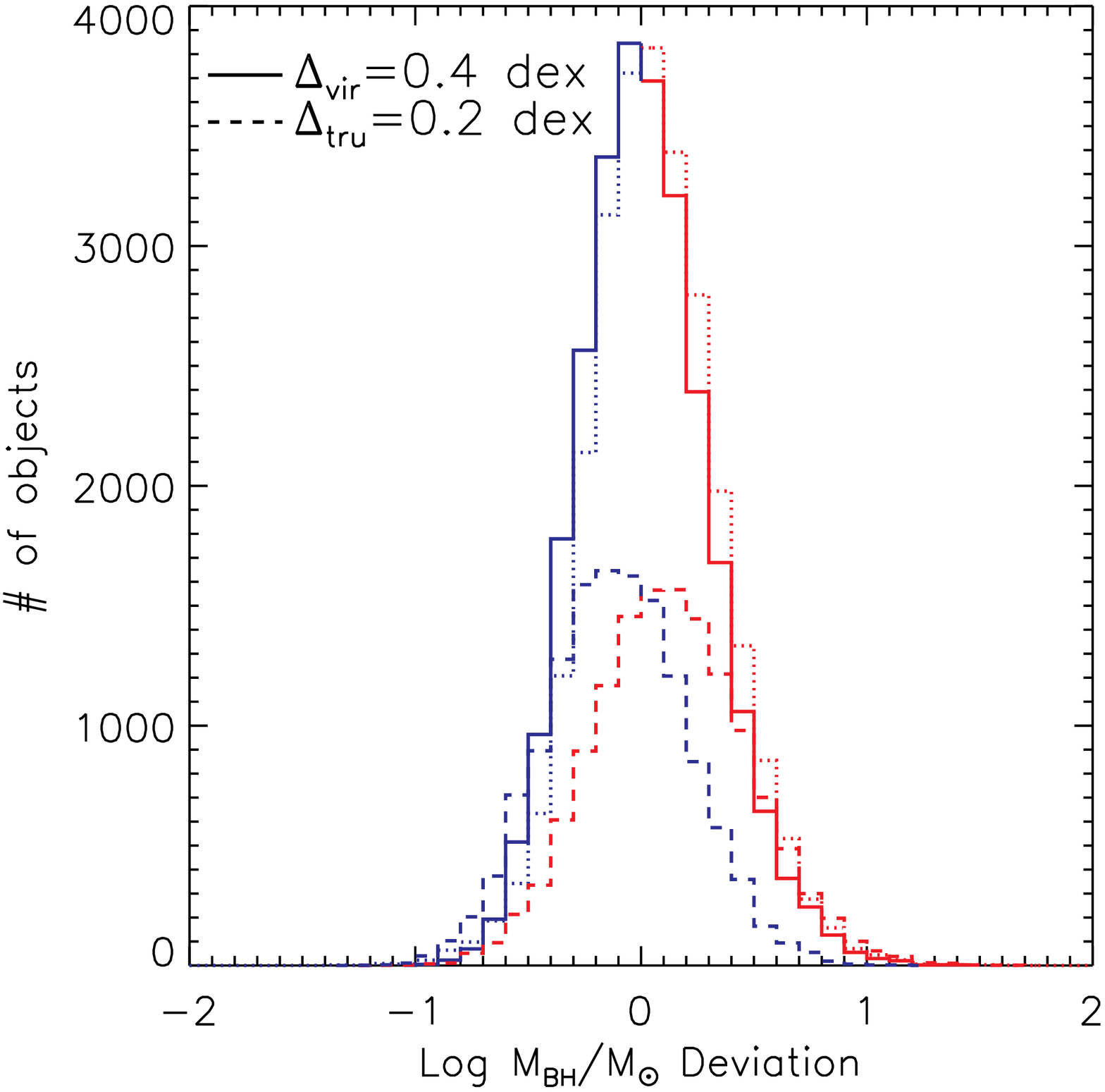}
    \caption{Simulated virial ({\em left}) and true ({\em middle}) BH masses. In addition to the
    Malmquist-type bias between the virial and the true BH masses (e.g., Shen et al. 2008b), the
    two samples divided using the median virial BH masses overlap substantially in their true masses.
    In the right panel we show the distributions of BH masses relative to the median values. The solid
    and dotted histograms show the distributions of deviation from the median value (which varies with redshift)
    for
    {\em observed} and {\em simulated} virial BH masses, which are
    in reasonable agreement. The dashed histograms show the distributions of {\em simulated} true
    BH masses of the two samples divided by the median {\em virial} BH masses. The median difference
    in the two samples are 0.4 dex in virial BH masses and 0.2 dex in true BH masses.}
    \label{fig:sim_BH}
\end{figure*}

Our simulated virial and true BH masses are shown in Fig.
\ref{fig:sim_BH}, where they are divided using the same definition
of our high/low mass subsamples in \S\ref{subsec:sample}. The
simulated virial BH masses reproduce the distribution of observed
virial masses well. The differences in the median values between
these two subsamples are $0.4$ dex and $0.2$ dex for the virial
and true BH masses respectively, i.e., there is not much
distinction between the true masses for the apparently high and
low mass populations. The dynamical range in the median $M_{\rm
halo}$ is then only $\sim 0.12$ dex given our assumed simple model
without the scatter $\sigma_{\rm BH}$. Take the typical host halo
mass $2\times 10^{12}\ h^{-1}M_\odot$, the expected difference in
bias for the {\em low} and {\em high} mass samples would be only
$\sim 5\%$, well below our $1\sigma$ uncertainties $\sim 10\%$.
Thus it is not surprising that we do not observe an appreciable
difference in the clustering strength of the {\em high} and {\em
low} mass quasar samples, given our current precision of the
clustering measurements. As in the luminosity dependence case,
when the scatter $\sigma_{\rm BH}$ is included, the expected
dependence could be even weaker.

\subsection{Color and Radio Dependences}\label{subsec:color_radio_dep}
Our clustering measurements of reddened, red and blue quasar
populations show consistent results, with differences that are
undiscernible given our uncertainty levels. A weak dependence on
color is expected if the color of a quasar is determined by the
physical conditions in the host galaxy of the SMBH more than by
its large-scale environment. Also, there is no strong correlation
of color with either luminosity or BH mass, indicating that this
diversity in quasar color occurs at all mass scales that our
sample probes. These reddened quasars are most likely
dust-reddened in the immediate nuclear environs and/or host galaxy
of the quasar, as argued by Richards et al. (2003) and Hopkins et
al. (2004), and there is a strong connection between
dust-reddening and broad absorption line quasars (BALQSOs; e.g.,
Reichard et al. 2003). Reassuringly, there seems to be no
appreciable difference between the clustering strength of BALQSOs
and regular quasars either (Shen et al. 2008a).

The most significant difference in clustering strength we have
found is for the FIRST-detected quasar sample. The relative bias
factor of radio-detected to radio-undetected quasars is $\sim
1.6$, using the correlation length values in Table 1. This implies
that, on average, radio-detected quasars live in more massive
halos than radio-undetected ones. Using the formula in Sheth et
al. (2001) for the halo bias factor, the estimated host halo mass
is $\sim 1-2\times 10^{13}\ h^{-1}M_\odot$ for FIRST-detected
quasars, and $\sim 1.5-3\times 10^{12}\ h^{-1}M_\odot$ for
FIRST-undetected quasars, at median redshift $\bar{z}\sim 1.4$. If
one uses the Jing (1998) fitting formula for the halo bias factor
instead (which gives the largest bias factor for fixed halo mass
compared with others), the resulting halo masses are $\sim 5\times
10^{12}-1\times 10^{13}\ h^{-1}M_\odot$ and $\sim 1-2\times
10^{12}\ h^{-1}M_\odot$ for FIRST-detected and undetected quasars
respectively. Accurate determinations of the host halo masses for
the radio-loud and radio-quiet populations again require better
clustering measurements and robust halo bias factor from
simulations. The trend observed in this study is consistent with
low redshift ($z<0.3$) observations by Mandelbaum et al. (2008)
based on clustering measurements (also see Lin et al. 2008 and
Wake et al. 2008) and lensing maps of radio-loud type 2 AGN. Both
studies find that for comparable virial BH masses (or galaxy
properties), radio quasars/AGNs reside in more massive halos than
their regular counterparts, indicating that radio quasars/AGN may
follow a different BH-halo relation from the normal population.


\section{CONCLUSIONS}
We have performed clustering measurements of quasars from a
uniformly selected sample in the fifth data release of SDSS, and
focussed on the dependences of quasar clustering on luminosity,
virial BH mass, quasar color and radio loudness.

Our main findings are the following:
\begin{enumerate}

\item[(i)] Strong luminosity dependence of quasar clustering at
$z<2.5$ is {\em not} detected for typical quasar luminosities.
While this is consistent with several theoretical predictions, we
caution that the dynamical range in luminosity probed in our
sample is narrow and our current sample size (especially when
split in luminosity/redshift) is not yet large enough to yield
accurate clustering measurements. However, the $10\%$ most
luminous quasars in our sample are more strongly clustered, which
is likely caused by the fact that halo bias increases rapidly with
mass at the high mass end. The final data release of SDSS-II (DR7,
Abazajian et al. 2008) will double the spectroscopic quasar sample
size and boost the clustering signal by a factor of $\sim 4$, thus
providing superior results over the current study.

\item[(ii)] Our measurements show weak or no dependence of quasar
clustering on virial BH mass at $z<2.5$. This is at least partly
due to the fact that virial masses are scattered around true BH
masses, where only the latter are presumably correlated with host
halo masses. The limited dynamical range in BH mass and our
limited sample size are other barriers to detecting this mass
dependence. However, even if we have accurate BH mass estimates,
the dependence of quasar clustering on BH mass may still be weak
if there is substantial scatter around the relation between BH
mass and halo mass. Thus the combination of precise BH mass
estimates (i.e., improved virial or new BH mass estimators) and
high-quality clustering measurements in future surveys can be used
to test how SMBHs grow in dark matter halos in the hierarchial
structure formation framework.

\item[(iii)] Blue, red and reddened quasars have similar
clustering strengths, at least indistinguishable at the current
levels of measurement uncertainty and dynamical range probed by
our sample.

\item[(iv)] Radio-loud quasars are more strongly clustered than
are radio-quiet quasars. This implies that radio-loud quasars live
in more massive dark matter halos and denser environments than
radio-quiet quasars, consistent with local $z<0.3$ observations
for radio-loud type 2 AGN (Mandelbaum et al. 2008) and radio
galaxies (Lin et al. 2008; Wake et al. 2008). Radio quasars
cluster more strongly than a regular quasar sample matched in
virial BH mass, indicating that the BH-halo connection might be
different for the radio population. Massive dark matter halos may
provide the necessary hot medium needed for radio activity.

\end{enumerate}

Future quasar clustering analyses should aim at both broader
dynamical ranges and larger sample sizes. Faint quasar detections
are the focus of many ongoing or planned wide-angle survey
projects such as BOSS (Schlegel et al.\ 2007) and Pan-STARRS
(Kaiser et al.\ 2002), which will both broaden the dynamical range
and enlarge the sample size. Because the halo bias factor
increases rapidly at the high-mass end, the clustering of the most
luminous and most massive quasars will provide the best test of
theoretical models (e.g., Lidz et al. 2006; Hopkins et al. 2007).
To obtain accurate measurements of quasar clustering at the high
luminosity end one needs a much larger control sample to cross
correlate with them; this control sample could be either a low
luminosity quasar sample or a high redshift galaxy sample.

\acknowledgements We thank the anonymous referee for helpful
comments. This work was partially supported by NSF grants
AST-0707266 (YS and MAS) and AST-0607634 (NPR, DVB, and DPS).

Funding for the SDSS and SDSS-II has been provided by the Alfred
P. Sloan Foundation, the Participating Institutions, the National
Science Foundation, the U.S. Department of Energy, the National
Aeronautics and Space Administration, the Japanese Monbukagakusho,
the Max Planck Society, and the Higher Education Funding Council
for England. The SDSS Web Site is http://www.sdss.org/.

The SDSS is managed by the Astrophysical Research Consortium for
the Participating Institutions. The Participating Institutions are
the American Museum of Natural History, Astrophysical Institute
Potsdam, University of Basel, University of Cambridge, Case
Western Reserve University, University of Chicago, Drexel
University, Fermilab, the Institute for Advanced Study, the Japan
Participation Group, Johns Hopkins University, the Joint Institute
for Nuclear Astrophysics, the Kavli Institute for Particle
Astrophysics and Cosmology, the Korean Scientist Group, the
Chinese Academy of Sciences (LAMOST), Los Alamos National
Laboratory, the Max-Planck-Institute for Astronomy (MPIA), the
Max-Planck-Institute for Astrophysics (MPA), New Mexico State
University, Ohio State University, University of Pittsburgh,
University of Portsmouth, Princeton University, the United States
Naval Observatory, and the University of Washington.

Facilities: Sloan

\clearpage


\end{document}